\documentclass[onecolumn]{aastex63}
\usepackage[english]{babel}

\usepackage{graphicx}
\usepackage{tabularx}
\usepackage{amsmath,amstext}
\usepackage{array}
\usepackage{longtable}
\usepackage[para]{footmisc}

\begin{document}
\defcitealias{Paper1}{K24a}
\defcitealias{Paper2}{K24b}
\defcitealias{Paper3}{R24}
\defcitealias{Readhead_1994}{R94}
\defcitealias{Wilkinson_1994}{W94}
\defcitealias{1996ApJ...460..612R}{R96}
\defcitealias{PR}{PR}
\defcitealias{CJ1}{CJ1}
\defcitealias{PWa}{PWa}
\defcitealias{PWb}{PWb}

\title{Exploring Compact Symmetric Objects with Complex Morphologies}
\author{E. E. Sheldahl}
\thanks{Grote Reber Doctoral Fellow at the National Radio Astronomy Observatory}
\affiliation{Department of Physics and Astronomy, University of New Mexico, Albuquerque, NM 87131, USA}
\author{G. B. Taylor}
\affiliation{Department of Physics and Astronomy, University of New Mexico, Albuquerque, NM 87131, USA}
\author{S. E. Tremblay}
\affiliation{National Radio Astronomy Observatory, Socorro, NM 87801, USA}
\author{W. Peters}
\affiliation{U.S. Naval Research Laboratory, 4555 Overlook Ave. SW, Washington, DC 20375, USA}
\author{S. Kiehlmann}
\affiliation{Institute of Astrophysics, Foundation for Research and Technology-Hellas, GR-71110 Heraklion, Greece}
\author{R. D. Blandford}
\affiliation{Kavli Institute for Particle Astrophysics and Cosmology (KIPAC), Stanford University, Stanford, CA 94305, USA}
\author{M. L. Lister}
\affiliation{Department of Physics and Astronomy, Purdue University, 525 Northwestern Avenue, West Lafayette, IN 47907, USA}
\author{T. J. Pearson}
\affiliation{Owens Valley Radio Observatory, California Institute of Technology, Pasadena, CA 91125, USA}
\author{A. C. S. Readhead}
\affiliation{Owens Valley Radio Observatory, California Institute of Technology, Pasadena, CA 91125, USA}
\author{F. Schinzel}
\thanks{An Adjunct Professor at the University of New Mexico.}
\affiliation{National Radio Astronomy Observatory, Socorro, NM 87801, USA}
\author{A. Siemiginowska}
\affiliation{Center for Astrophysics—Harvard and Smithsonian, 60 Garden St., Cambridge, MA 02138, USA}
\author{R. Skalidis}
\affiliation{Owens Valley Radio Observatory, California Institute of Technology, Pasadena, CA 91125, USA}

\begin{abstract}
Compact symmetric objects (CSOs) are a unique class of jetted active galactic nuclei (AGN) defined by sub-kpc radio emission, showing radio structure on both sides of the central engine. CSOs tend to exhibit little to no relativistic beaming, thereby allowing us to determine their physical characteristics, such as the magnetic field strength and particle energy density. Selected with a literature search, we describe VLBI observations, imaging, and analyses of 167 CSO candidates. We identified 65 new bona fide CSOs, thus almost doubling the number of known bona fide CSOs to 144. With our greater breadth of sources, we confirm that edge-dimmed CSOs (CSO-1s) may represent a more diverse population than originally expected. We highlight a number of CSOs with complex morphologies, including candidates for supermassive binary black holes (SBBHs) and CSOs that appear to have morphologies akin to wide-angle tail (WAT) galaxies, which could perhaps indicate that some CSOs are experiencing a galactic merger.
\end{abstract}

\section{Introduction} \label{intro}

In the unified theory of radio loud active galaxies \citep{1978Natur.276..768R, 1984RvMP...56..255B, 1984ApJ...278..499A, 1991ApJ...378...47M, Antonucci_1993, 1995PASP..107..803U, Netzer_2015, 2019ARA&A..57..467B}, the observed properties of galaxies strongly depend on the orientation of the jet axis to the line of sight. This theory has been highly successful in explaining many of the major differences between radio galaxies and quasars. The first multi-frequency ``hybrid maps" -- i.e., VLBI maps incorporating the closure phase -- showed that the radio loud active galactic nuclei (jetted AGN) with flat spectra consisted of a flat spectrum ``core" component at one end of a steep spectrum jet \citep{1978Natur.276..768R,1979ApJ...231..299R}. These are now commonly known as ``core-jet" sources. The most likely explanation for their asymmetric appearance is due to relativistic beaming of the jet towards the observer and the counter-jet away from the observer, resulting in flux variability on short timescales, an increase in flux from the incident jet, and massively decreased flux from the counter jet (\citealt{Rees_1966}, \citealt{2019ARA&A..57..467B}). Other possible explanations are complex environmental interactions suppressing the flux of one of the jets or uneven fueling of the central black hole. \par

Other important subsets of jetted AGN include the objects that are dominated by compact components in the vicinity of their nuclei, amongst which there are three major radio spectral classes: (i) compact flat spectrum; (ii) compact steep spectrum (CSS); and (iii) peaked spectrum (PS) objects. These have recently been thoroughly reviewed by \citet{O'Dea_2021}. Although spectral classification has been a powerful tool in the study of jetted AGN \citep{O'Dea_1998}, \citet{2024arXiv240813077D} have recently demonstrated that taking subsets of these three classes, as is often done in jetted AGN studies, can lead to significant fractions of the population under study being excluded. In order to overcome the difficulties of possible biases in the radio-spectrum selection approach, \citet{Wilkinson_1994} introduced the morphologically-based classification of compact symmetric objects (CSOs). Their definition of CSOs was twofold: sources that possess (1) symmetrical lobe emission either about an observed central core or with sufficient evidence of one if none is detected, such as compressed lobe edges, and (2) total radio emission spanning less than 1 kpc. \par

Unfortunately, sometimes in classifying compact jetted AGN as CSOs, morphological or spectral classification has been overly relied upon and aspects like variability and beaming have been ignored. This has led to the mis-classification of some core-jets, blazars, and other non-qualifying objects as CSOs or CSO candidates in the literature. For example, the blazar PKS 1413+135 was originally mistakenly classified as a CSO due to its apparent symmetric lobe emission but was later refuted by analyzing its variability \citep{Readhead_2021}. To address this issue, a comprehensive review of CSOs was carried out by \citet{Paper1}, \citet{Paper2}, and \citet{Paper3}, hereafter \citetalias{Paper1}, \citetalias{Paper2}, and \citetalias{Paper3}. \par

These authors proposed that in addition to the criteria spelled out in \citet{Wilkinson_1994}, bona fide CSOs should not exhibit (1) fractional flux density variability greater than $\sim$20\% per year, or (2) apparent superluminal motion (v$_{app}$) in excess of 2.5 c. In a comprehensive search of the literature, \citetalias{Paper1} identified 79 bona fide CSOs and 167 class A-candidate CSOs for immediate follow-up with Very Long Baseline Array (VLBA) observations. In this paper, we describe our analysis of these 167 candidates with the VLBA and Very Large Array (VLA) and how we identified 65 of them as bona fide CSOs, almost doubling the known population. \par

\citet{Tremblay_2016} discovered that there are two major types of CSOs: an edge-dimmed class designated ``CSO-1s" by \citetalias{Paper1}, and an edge-brightened class designated ``CSO-2s" by \citetalias{Paper1}. \citetalias{Paper3} divided CSO-2s into three subclasses: 2.0s, which have outer lobes with prominent hotspots, 2.2s, which have lobes that do not have prominent hotspots and are extended perpendicular to the jet axis, and 2.1s, an intermediate category between 2.0s and 2.2s. \par
Some previous observing campaigns looking for CSOs incorporate the study of supermassive binary black holes (SBBHs) \citep{Tremblay_2016}. In fact, the CSO J0405+3803 was sufficiently resolved to be able to confirm it as a parsec-scale SBBH using VLBI observations \citep{Bansal_2017}. SBBHs start when two galaxies, each with a central supermassive black hole (SMBH), merge. Through complex interactions involving dynamical friction, the SMBHs are driven closer and closer to each other until they eventually reach a phase where they orbit a few parsecs from each other, at which point they are officially a binary system \citep{Agazie_2023}. How SBBHs go from this stage of $\sim$parsec-scale separation to merging is still a mystery since at these separations, final coalescence driven purely by gravitational waves would take longer than the Hubble time \citep{Agazie_2023}, though there are many theories to how this ``final-parsec problem" could be solved (\citealt{Berczik_2006}, \citealt{Holley-Bockelmann_2015}) and also evidence that these mergers are relatively common \citep{Haehnelt_2002}. Since the process of two black holes merging necessitates them orbiting within parsecs of each other for long periods of time, we expect to find a few while studying parsec-scale radio galaxies. Follow-up observations looking to assess the kinematics of potential orbiting SMBHs can also be used to measure CSO jet expansion speeds and ages. It has been shown that because of their small linear sizes, we can notice morphological changes in CSOs after only a few years (e.g. \citealt{Gugliucci_2005}, \citealt{Bansal_2017}). Therefore, we take note of CSOs that demonstrate evidence of being SBBHs and mark them as candidates worth further investigation. \par
This paper is structured as follows: In Section \ref{methods}, we explain our observation setup, data calibration, and image making methods of our 167 A-candidate sources. Section \ref{analysis} explains how we sorted the bona fide CSOs from the A-candidates and what parameters we extracted from them for further analysis. Section \ref{statsofcsos} contains our statistical analyses of the extracted CSO parameters. In Sections \ref{htcsos}-\ref{highfluxratio}, we detail a number of sources worthy of deeper future analysis: CSOs that resemble wide-angle tail (WAT) galaxies (Section \ref{htcsos}), medium symmetric objects (MSOs, Section \ref{msos}), candidates for supermassive binary black holes (SBBHs, Section \ref{sbbhcands}), and CSOs with a large ratio between the flux of their radio lobes (Section \ref{highfluxratio}). Our summary and conclusions are in Section \ref{summary}. Descriptions of a few individual sources with interesting characteristics are in Appendix \ref{indivsources}. Tables containing data on our newly observed bona fide CSOs, rejected A-candidates, and indeterminate A-candidates are in Sections \ref{bonafidesection}, \ref{rejectedsection}, and \ref{indetsection}, respectively. \par

In this paper, we define the spectral index $\alpha$ by $S \propto \nu^{\alpha}$ and adopt a cosmology with $H_0 = 71$ km s$^{-1}$ Mpc$^{-1}$, $\Omega_m = 0.27$, and $\Omega_\Lambda = 0.73$.
\vfill
\section{Observations \& Methods} \label{methods}
\subsection{Sample Selection}
To compile our list of candidate CSOs worth further investigation, we followed the methods described in \citetalias{Paper1}, which we briefly summarize here. They used the Astrophysics Data System (ADS\footnote{https://ui.adsabs.harvard.edu/}) to search for mentions in the literature for CSOs, GPSs, and CSSs, compiling a list of 3175 objects that were claimed to be CSOs or CSO candidates. By investigating images and spectra of each source, they sorted them into four categories:
\vskip 6pt
1. Bona fide CSOs, for which there was available evidence to confidently say they did not violate any of the four defining criteria
\vskip 6pt
2. A-candidates, for which the available evidence did not conclusively confirm or rule out the source as a bona fide CSO
\vskip 6pt
3. B-candidates, which lacked sufficient evidence to make a proper judgment call
\vskip 6pt
4. Rejected candidates, which violated one or more of the criteria
\vskip 6pt

They verified 79 sources as bona fide CSOs and selected 167 A-candidates. We were awarded VLBA and VLA time to conduct follow-up observations of these candidates.

\subsection{VLBA Observations}
Our objects were observed with the VLBA under project code BT152, which consisted of seven epochs, each approximately 24 hours in length, in three frequency bands: 4.612-5.124 GHz, 8.11225-8.62425 GHz, and 14.91175-15.42375 GHz (hereafter 5 GHz, 8 GHz, and 15 GHz). There were four intermediate frequencies (IFs) per band, each covering 128 MHz. In order to obtain roughly equal image sensitivity in each frequency band, we observed each target for average times of 196, 504, and 1386 seconds at 5 GHz, 8 GHz, and 15 GHz, respectively. We did not phase reference our targets. In total, we observed 171 sources: the 167 A-candidates from \citetalias{Paper1} plus three bona fides (J0909+1928, J1025+1022, and J1205+2031) and one rejected source (J0048+3157) that had been reclassified by the time \citetalias{Paper1} was published. Each source was observed during only one epoch. Further details of the observations are given in Table \ref{vlbaobsdetails}. For a list of all sources observed, see Appendices \ref{bonafidesection}, \ref{rejectedsection}, and \ref{indetsection}. \par
Note that J0909+1928, J1025+1022, and J1205+2031 are included in the 79 CSOs of \citetalias{Paper1}, but we include them in this analysis since these maps have not been previously published elsewhere. In future sections when we compare the 79 CSOs compiled in \citetalias{Paper1} and the ones discussed in this paper, we consider J0909+1928, J1025+1022, and J1205+2031 to be \citetalias{Paper1} CSOs, but we update them with new data from this analysis. We use the phrase ``newly confirmed CSOs" to refer to the 65 A-candidates that we classified as bona fide CSOs, while the term ``newly observed CSOs" refers to J0909+1928, J1025+1022, and J1205+2031, plus the 65 CSOs verified from the A-candidates, totaling 68 objects. \par

\begin{table}[h]
    \centering
    \caption{Details of the VLBA observations. Col (1): Observing epoch label. Col (2): UT date start of observations. Col (3): Number of target sources (excludes calibrators). Col (4): Number of antennas. Col (5): Average RMS noise for the final images of all the bona fide CSOs at each frequency.}
    \begin{tabular}{c|cccc}
        \hline \hline
         Epoch & Obs. Start & \parbox[c]{0.5in}{\centering\# of Targets} & \parbox[c]{0.45in}{\centering Antennas Active} & \parbox[c]{1.1in}{\centering RMS Noise at 5/8/15 GHz (${\rm \mu Jy\,beam}^{-1}$)}\\
         (1) & (2) & (3) & (4) &(5)\\\hline
         A & 2021 Oct 07 & 24 & 10\tablenotemark{a} & 313/182/152\\
         B & 2021 Oct 08 & 25 & 9 & 174/138/122\\
         C & 2021 Oct 31 & 25 & 9 & 139/112/108\\
         D & 2022 Jan 23 & 25 & 8 & 158/135/98\\
         E & 2022 May 26 & 24 & 8 & 100/125/90\\
         F & 2022 Jun 18 & 24 & 9\tablenotemark{b} & 245/173/133\\
         G & 2022 Jun 19 & 24 & 9\tablenotemark{b} & 153/130/112\\\hline
    \end{tabular}
    \tablenotetext{a}{Only 9 antennas at 5 GHz}
    \tablenotetext{b}{Only 8 antennas at 8 GHz}
    \label{vlbaobsdetails}
\end{table}

\subsubsection{Calibration}
We calibrated our data manually using the Astronomical Image Processing System (\textsc{AIPS}) \citep{Greisen_2003}. In addition to the target source, each epoch included scans of the calibrator sources J0927+3902, J1146+3958, J1256-0547 (3C 279), J1310+3220, and J1407+2827 (OQ +208). We first applied parallactic angle corrections using CLCOR and amplitude corrections using ACCOR and APCAL. We then calibrated the phases using FRING, using a single scan on a bright calibrator source with as many active antennas as possible and reasonably level gains on all antennas. We then applied bandpass calibration with BPASS, using the same calibrator source as specified in FRING. After applying all the calibration tables, we split out individual calibrated UVFITS files of each source with the frequency channels in each IF averaged together, ready for imaging.

\subsubsection{Image Making}
Self-calibration and imaging of the sources were performed in Difmap \citep{Shepherd_1994}. Despite the large number of sources and necessity of making three images per source, imaging was performed manually to ensure each source received optimal treatment. Since CSO classification is reliant on properly determined morphology and our sources covered a wide range of shapes, flux densities, and sky coordinates, we wanted to avoid the potential pitfalls of applying an automated imaging algorithm. The general process for imaging started with checking for and flagging (ignoring) bad data using RADPL and VPL, such as scans with low SNR and data points with excessively low or high amplitudes. We then initialized the model with a 1 Jy point source at the phase center using STARTMOD. This step was necessary because we did not do phase referencing. As a result, absolute position information was lost, requiring us to focus the image at the phase center manually. Next, we compiled our model image through a two-part process of drawing clean boxes around significant emission and running CLEAN and SELFCAL to perform phase-only self-calibration, repeating this process until the residual image was mostly noise-like. We then switched from uniform to natural weighting and began amplitude self-calibration by using GSCALE TRUE to apply a per-antenna amplitude correction. After that, we remade the model with our new calibration applied before applying more amplitude self-calibration using SELFCAL TRUE,TRUE with a solution interval equal to the scan length. The pixel sizes of the Stokes I maps were 0.25 milliarcseconds (mas) at 5 GHz, 0.2 mas at 8 GHz, and 0.1 mas at 15 GHz unless otherwise stated. This corresponded to approximately 7.6, 5.5, and 5.7 pixels across the minimum full width half maximum (FWHM) restoring beam dimension in the naturally-weighted images at 5, 8, and 15 GHz, respectively. In a few instances where maps needed to be extremely large to encompass all of the extended emission, we increased the pixel size so as to not exceed the maximum image pixel dimensions imposed by \textsc{AIPS}. The contour levels are spaced apart by factors of 2, with the lowest contour level set at three times the RMS noise. \par

Because absolute position information was lost through the phase self-calibration step, sometimes sources appeared to be misaligned at different frequencies. This manifested as noticeable stripes of inverted to steep spectrum (blue to red) on the eventual spectral index maps. In these cases we applied the 8 GHz model of the source to our 5 and 15 GHz images, ran one iteration of phase-only self-calibration, then cleared the model and restarted the image process with these additional calibration solutions. This brought source components into better alignment with the 8 GHz images. This method does not account for frequency-dependent core shift. In a sample of 29 AGN, \citet{Kovalev_2008} found the maximum core shift to be 1.4 mas between 2.3 and 8.6 GHz, with a median shift of 0.44 mas. It is possible that with a more rigorous analysis considering core shift, some spectral index maps would exhibit slightly less striping, leading to cleaner maps. \par

\begin{figure}[h]
    \centering
    \includegraphics[page=1,trim={5.1cm 2.8cm 5.1cm 4cm},clip,width=0.68\linewidth]{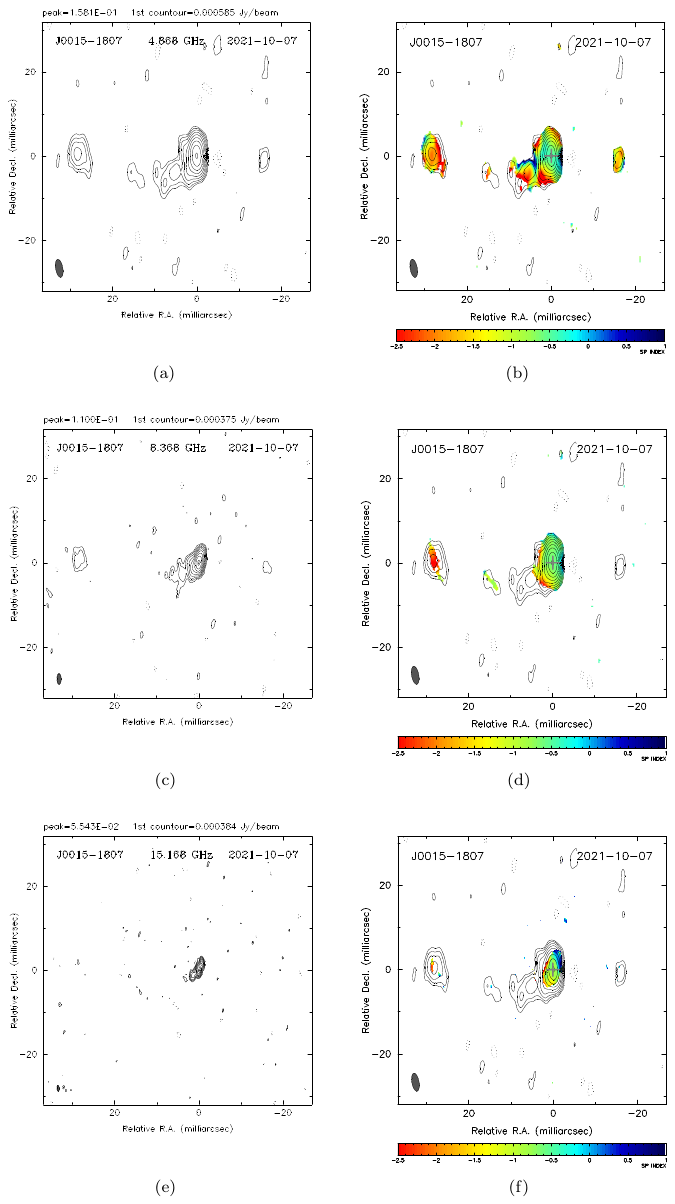}
    \caption{Example of the six-image compilation we made for each source. Subfigures a, c, and e are 5, 8, and 15 GHz Stokes I images, respectively, with the contour level starting at three times the RMS noise in each. Subfigures b, d, and f are the 5-8, 5-15, and 8-15 GHz false color spectral index maps, respectively, with 5 GHz contours overlaid starting at three times the RMS noise level and crosses identifying the core if we detect it. Core locations were determined through a combination of fitting model components and placement by eye and are merely meant to guide the reader's eye. Quantitative core locations are available upon request.}
    \label{allmap_example}
\end{figure}

We made three spectral index maps for each source: 5-8, 5-15, and 8-15 GHz. First, we reconvolved the beam shape and modified the pixel size and map size of the higher frequency image to match that of the lower frequency image. The FITS files were then loaded into \textsc{AIPS} and the RMS noise was measured by taking the median RMS among rectangular regions located in the four corners of each image. We used two times the RMS in the image as a blanking threshold; emission below this level was excluded from the maps. We then combined the images into a single map using the \textsc{AIPS} task COMB. The maps were overlaid with 5 GHz contours using MAPPLOT\footnote{\url{https://sites.astro.caltech.edu/~tjp/citvlb/vlbhelp/mapplot.html}}.
We displayed a range of $-$2.5 to 1 in spectral index; this covers the steepest negative spectral index due to synchrotron self-absorption \citep{1968ARA&A...6..321S} and a sufficient range of positive spectral index to include inverted core components while avoiding most image artifacts. See Figure \ref{allmap_example} for an example of the images we made for each source. \par

\subsection{VLA Observations}
We supplemented our VLBA observations with archival observations from the VLA. These data came from two origins: our previous campaign to assess CSO candidates as potential phase and polarization-leakage calibrators, detailed in EVLA Memo \#224\footnote{\url{https://library.nrao.edu/public/memos/evla/EVLAM_224.pdf}}, and the VLA Low-band Ionosphere and Transient Experiment (VLITE, \citealt{Clarke_2016}). The data from our previous campaign were observed as part of the TPOL0003 project code, which is also used for monitoring polarization calibrators. They spanned 1-15 GHz split into four frequency bands, hereafter referred to as 1.5, 5.5, 9, and 14 GHz. The data were observed in the BnA $\rightarrow$ A configuration for seven of the eight epochs and in the A configuration for the remaining epoch. They consisted of all but four of our A-candidates; not included are J0037-2145, J0301+3512, J0347+2004, and J1326+5712. This is because our A-candidates list had not been finalized by the time our VLA data were collected. The VLA observations clearly measured all of the available flux in the compact pc-scale structure. \par
VLITE records data during nearly all regular VLA operations that is calibrated, imaged, and cataloged using pipeline processing to create an archive of source measurements \citep{Polisensky_2016}. We obtained VLITE data from approximately 321.7-355.3 MHz with a central frequency of about 338.5 MHz (hereafter referred to as 340 MHz). VLITE flux densities for the CSOs from data taken during the TPOL0003 project were available for all sources except J0429+331. Although we did not expect much variability in CSOs, these simultaneous measurements allowed us to constrain the low frequency spectral shape without any temporal uncertainty.
\subsubsection{Calibration}
For this analysis, we used the Common Astronomy Software Applications (\textsc{CASA}, \citealt{CASA}) package version 6.6.3.22, but instead of using the VLA PI Pipeline described in the original memo, we recalibrated our data manually, generally following the \textsc{CASA} Guides VLA Continuum Tutorial for 3C391\footnote{\url{https://casaguides.nrao.edu/index.php?title=VLA\_Continuum\_Tutorial\_3C391-CASA6.4.1}}. \par

This calibration was effective for unresolved point sources, which is what we generally expect from CSOs at the arcsec-scale resolution of the VLA. However, one source, J0552-0727, had a triple structure at 5.5 GHz in which the secondary components were bright enough to render the point source model unsuitable. Because it has a similar structure at 9 GHz, we used a 9 GHz model image to simulate a 5.5 GHz measurement set with the proper configuration, sky coordinates, and on-source time using the \textsc{CASA} task \textit{simobserve()}. After making an image with the new simulated measurement set, we applied this model to the empirical 5.5 GHz measurement set before the gain calibration step. This allowed us to successfully obtain calibrated data for the source at 5.5 GHz. \par
The VLITE data were separately calibrated using their dedicated Astrophysics Data Processing Pipeline\footnote{For more information, see \url{http://vlite.nrao.edu/imaging.shtml}}.

\subsubsection{Image Making}
With our VLA data calibrated, we made several Stokes I images for each bona fide CSO. These consisted of one for each of the sixteen spectral windows per frequency band, totaling 64 images, one full-bandwidth image for each frequency band, and 8 GHz images with a bandwidth matching that of our VLBA observations (8.11225-8.6242 GHz) to compare their flux densities. In this way we made 69 VLA images for each CSO. 
Due to the three-component structure of J0552-0727, in order to measure the spectral index of each component separately in our analysis, we made individual images in each spectral window in \textsc{CASA} using the same beam as the lowest frequency good quality image. Unfortunately, our 1.5 GHz observations were only barely resolved, which prevented us from distinguishing the individual components for spectral analysis, so the lowest frequency image of good quality was at 4.935 GHz.

\subsection{Variability and Kinematic Measurements}
Our observations did not include variability or kinematic measurements, not allowing us to directly address the two criteria added by \citetalias{Paper1} when vetting A-candidates. In \citetalias{Paper1}, some CSOs had associated variability data as part of the OVRO 40m Telescope Monitoring Program at the Owens Valley Radio Observatory (OVRO; \citealt{Richards_2011}). Only a handful of bona fide CSOs had variability data (J0428+3259, J1537+8154, J1823+7938, J1855+3742, J1935+8130, and J2153+1741), and did not have the same time or frequency coverage, meaning our data from source to source were inconsistent. Because we did not have uniform data for all of our sources, it is outside the scope of this paper to classify CSOs directly based on the variability criterion. To address the lack of variability data, we are in the process of adding bona fide CSOs to the OVRO monitoring program. \par
As for kinematic measurements, we lacked multiple observations of the 167 A-candidates far enough separated in time to make adequate estimates on the apparent superluminal motion of the jets. However, this is an avenue we are very interested in pursuing with future VLBA follow-up. Observations of our sources spanning about five years would be sufficient, such as in \citet{Gugliucci_2005}. As such, we did not address the apparent superluminal motion criterion when classifying our sources.

\section{Analysis} \label{analysis} 
The primary goal of our VLBA observing campaign was to use spectral index maps to classify the 167 class A-candidates of \citetalias{Paper1}. In all, we newly confirmed 65 sources as bona fide CSOs. We applied the following four criteria for confirming an object as a bona fide CSO:
\begin{enumerate}
    \item Largest size of the radio emission $<$ 1 kpc
    \item Emission on both sides around a center of activity (core). A source can still satisfy this criterion even if the core is not detected if it has typical features, such as lobes or hotspots straddling the putative core position
    \item Fractional flux density variability $<$ 20\%$\rm \,year^{-1}$
    \item Apparent superluminal motion (v$_{app}$) $<$ 2.5 c
\end{enumerate}
Our observations did not include variability or kinematic measurements, but at the time the VLBA observations were made, there were not any literature data to our knowledge that showed any target sources conflicting with criteria 3 and 4, otherwise we would not have selected them as A-candidates. We therefore mainly used criteria 1 and 2 to identify bona fide CSOs. To accomplish this, we interpreted compact regions of flat or inverted spectra as core components and steep spectrum regions as jet or lobe components. We looked for two jets and/or lobes straddling a core or situated such that it would be reasonable to expect an undetected core between them. See Table \ref{Acandpercentage} for our final numbers after classification.

\subsection{Identifying Bona Fide CSOs and Sub-Classifying CSOs}

Great care is required both in classifying jetted AGN as bona fide CSOs (\citetalias{Paper1}) and in determining which of the subclasses (CSO-1, CSO-2.0, CSO-2.1, and CSO-2.2) a CSO belongs to. The procedure adopted by \citetalias{Paper1} for identifying bona fide CSOs was that the sources were first triaged by subsets of the authors to filter out obvious core-jet objects, based on morphology, spectral index, apparent speed of components, and/or variability. Any difficult cases were then discussed and decided upon by all co-authors. Any cases that could not definitively be decided at that point were tagged for follow-up observations. The procedure adopted by \citetalias{Paper3} for determining the sub-classification of the CSOs was that the sources were subject to a blind test of selecting the CSO subclass in ignorance of the redshift, size, and luminosity of the objects. This was done in order to not introduce any unconscious biases into the classifications.\par

In this study, the same level of care was applied in selecting the bona fide CSOs as in \citetalias{Paper1}, and blind tests were carried out by three of us (ES, GT, and ST) in determining the CSO subclasses. Any difficult cases were then subject to detailed discussion between these three co-authors. We are confident, therefore, that the classifications we present in this paper can be relied upon, and that we have not misclassified any of the objects.

\subsection{Reporting CSO Parameters} \label{csoparams}
After selecting our bona fide CSOs, we extracted a series of parameters from them for statistical analyses. A full list of the parameters extrapolated from our images can be found in Appendix \ref{bonafidesection}. In this section, we describe in detail how each parameter was measured. \par
Spectroscopic redshifts were gleaned from the literature and the NASA/IPAC Extragalactic Database (NED). The angular size of each source was measured using an on-screen pixel ruler at the largest separation between the second lowest contours of the source structure, using the lowest frequency map in which the source was resolved. This was the 5 GHz map for all but one source (J0552-0727), for which we used the VLA 9 GHz map instead (for our exact method, see Appendix, Section \ref{j0552indiv}). The lowest contour was set at three times the RMS noise level. We estimated the size measurement uncertainty to be 1.5 times the FWHM diameter of the beam measured along the source axis. \par

We only calculated a linear size for sources with spectroscopic redshift information. For these we applied the 1 kpc cutoff criterion. The linear size was calculated by computing the angular diameter distance, which, from \citet{Hogg_1999}, is
\begin{equation}
    D_A = \frac{c}{H_0(1+z)} \int_{0}^{z} \frac{dz^{\prime}}{\sqrt{\Omega_m (1+z^{\prime})^3 + \Omega_\Lambda}}
    \label{angdist}
\end{equation}
where c is the speed of light, z is the redshift, and the cosmological parameters are given in Section \ref{intro}. Uncertainties were propagated from the angular size measurement uncertainties. \par

The turnover frequencies and peak flux densities were extracted from the VLA observations plus VLITE, which gave us valuable data at 340 MHz. Uncertainties were calculated by adding the RMS noise and a certain percentage of the flux density (15\% for VLITE, 5\% for 1-15 GHz) in quadrature. The RMS noise was measured for the 1-15 GHz VLA data using the CASA task IMFIT and for the VLITE data by using Python Blob Detector and Source Finder (PyBDSF). 15\% is the standard uncertainty for VLITE \citep{Polisensky_2016} and 5\% was chosen for 1-15 GHz based off of the upper bound of the expected flux scale uncertainty in \citet{Perley_2017}. Using these data, we fitted a spectral index curve for each bona fide CSO using the Scipy \citep{Virtanen_2020} function \textit{curve\_fit}. The formula used with \textit{curve\_fit} was

\begin{equation}
        S = S_{ref}(\nu/\nu_{ref})^{\alpha+\beta(\log_{10}(\nu/\nu_{ref}))}
    \label{spix_eq}
\end{equation}
where $S_{ref}$ is the flux density at the reference frequency $\nu_{ref}$, $\nu$ is the frequency, and $\alpha$ and $\beta$ are both coefficients in the spectral index term. We recorded the optimized parameters of each fit, which were $S_{ref}$, $\alpha$, and $\beta$, was well as calculated a coefficient of determination (R$^2$). \par

The VLITE data spanned $\sim$33.6 MHz total, compared to 1 GHz for our 1.5 GHz observations and 2 GHz each for our 5.5, 9, and 14 GHz observations. Therefore, even though it gave us valuable spectral information below 1 GHz, it was not enough to significantly alter the calculated spectral index in many cases. To avoid over-interpreting our data, we reported limits on the turnover frequency and flux density if the derived peak frequency within 1$\sigma$ did not overlap with our 0.3385-15 GHz range. In cases where this occurred, the peak frequency was projected to be below 0.3385 GHz. We set an upper limit on the frequency and a lower limit on the flux density at 0.3385 GHz if the VLITE data suggested an increasing flux density below 0.3385 GHz, and limits at 1 GHz if the VLITE data suggested there was a peak between 0.3385-1 GHz. We did not estimate a turnover if the R$^2$ value for our fit was less than 0.9. For some sources, some frequency bands appeared to be offset from a smooth spectral index curve. 
This might be attributed either to a spectral break at high frequencies or a calibration offset. In these cases, we noted which frequency bands were excluded from the fit. Uncertainties were calculated using the uncertainties of the optimized parameters given by \textit{curve\_fit}. \par

For J0552-0727, after finding the pixel coordinates representing the peak of each of its three components at 4.935 GHz, we measured the flux density of these pixels across all of our images. We used a different method for this source compared to the other sources because we did not want to introduce errors through estimating the integrated flux density of complex lobe structures. \par
Our method for determining the CSO class is taken from \citetalias{Paper3} and is described in in Section \ref{intro}. If we were unsure about a CSO class due to confusing maps, we left it unclassified. Core fractions were measured by modelfitting our final images in Difmap using MODELFIT and measuring the integrated flux density within a circular Gaussian placed at the position of the suspected core component, then dividing that Gaussian integrated flux density by the total integrated flux density of the source. We performed this step at 8 GHz only. \par

\subsection{Rejected \& Indeterminate Candidates}
Since this is a morphological study, the majority of our rejections were based on morphology. Three of our sources were rejected based on linear size $>$1 kpc: J1011+7124 (1.562 $\pm$ 0.030 kpc), J1052+3811 (1.561 $\pm$ 0.023 kpc), and J2137-2042 (1.245 $\pm$ 0.025 kpc). Due to their size, we investigated if any of these sources were medium symmetric objects (MSOs) in Section \ref{msos}. We rejected one source, J0048+3157, due to the variability criterion using 15 GHz light curve data from the 40m telescope at OVRO. See Appendix \ref{rejectedsection} for the full list of rejected sources. \par 
As would be expected with a sample of sources this large, some were more difficult to image and classify than others. The (u,v) coverage of sources with negative or near-zero declination was not as good as that of the other objects, and were prone to higher noise, lower brightness, and flattened beams. In addition, as detailed in Table \ref{vlbaobsdetails}, some observing epochs had fewer antennas than others, thereby degrading our (u,v) coverage. Many sources had such low brightness at higher frequencies that self calibration proved detrimental for resolving believable structure. We established a cutoff of 20 m${\rm Jy\,beam}^{-1}$ for the peak flux at 15 GHz, below which we largely removed the 15 GHz image from consideration in our source classifications. We classified any sources for which we were not confident of the morphology as indeterminate (Appendix \ref{indetsection}). \vspace{8mm}

\begin{table}[h]
    \centering
    \caption{Results of analysis of A-Candidates}
    \begin{tabular}{c|cc}
        \hline \hline
        Classification & Number & Percentage of Total \\\hline
        Bona fide CSO & 65 & 38.9\% \\
        Indeterminate & 46 & 27.5\% \\
        Rejected CSO & 56 & 33.5\% \\\hline
    \end{tabular}
    \label{Acandpercentage}
\end{table}

\section{Statistics of CSOs} \label{statsofcsos} 
\subsection{Redshift Distributions \& Core Fractions}
\begin{figure}[h]
    \centering
    \includegraphics[width=0.45\linewidth]{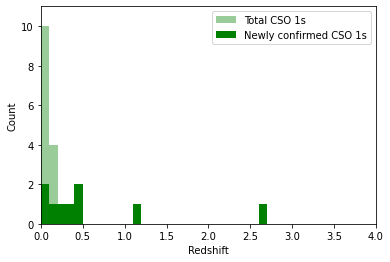} \hfill
    \includegraphics[width=0.45\linewidth]{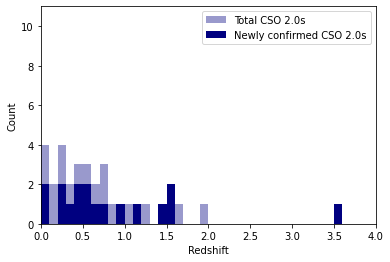} \\
    \includegraphics[width=0.45\linewidth]{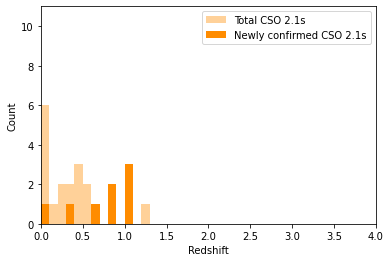} \hfill
    \includegraphics[width=0.45\linewidth]{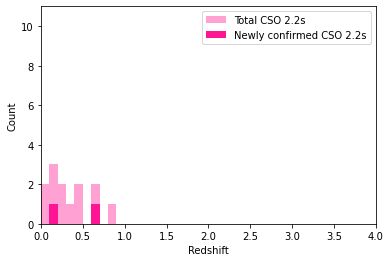}
    \caption{Spectroscopic redshift distributions of all confirmed CSOs, separated by morphological class. Newly confirmed CSOs are emphasized with dark colors in front of total distributions in light colors.}
    \label{zdist}
\end{figure}

As shown in Table \ref{zpercent}, we confirmed 65 new bona fide CSOs, adding on to the 79 previously verified ones from \citetalias{Paper1} and almost doubling the sample size. We also almost doubled the number of confirmed CSOs with known spectroscopic redshifts, adding 37 new ones to the previous 54. The redshift distributions of all verified CSOs, and CSO classifications are plotted in Figure \ref{zdist}, in which our newly confirmed CSOs are highlighted. CSO-1s are noticeably peaked at redshifts less than 0.1, which we expect because CSO-1s tend to be less luminous than other types. However, of the CSO-1s verified in \citetalias{Paper1}, eight have z $\leq$ 0.1 and the remaining three have z $\leq$ 0.2. This means that the low-redshift peak is largely because of \citetalias{Paper1} CSOs and our study introduces a whole slew of newly confirmed CSO-1s that have a more uniform redshift distribution extended out to at least z $\sim$ 2.7. There are two outliers among the CSO-1s with abnormally high redshifts, those being J1855+3742 (z = 1.12) and J1258+5421 (z = 2.65277). Previously, the highest redshift known CSO-1 was J0832+1832 with a redshift of 0.154, so it's an unanticipated result that we detect a more diverse sample of CSO-1s. \par
As for the other CSOs with known redshifts, 2.0s make up the largest group at 34, with 2.2s being the smallest at only 13. Groups with only a few CSOs are subject to small-numbers statistics, but the redshift distributions of CSO-2.0s and 2.2s are close to uniform with a slight peak near lower redshifts, likely due to bias from our image sensitivity. The maximum redshift at which we detect CSO-2.1s and 2.2s is less than for our 2.0s, possibly suggesting that 2.1s and 2.2s occur later in the CSO life cycle as touched on in \citetalias{Paper3}, but it is more likely that this is observational bias because these sources are generally more diffuse, meaning we would need greater brightness sensitivity to distinguish them at higher redshifts. \par

\begin{table}[h]
    \centering
    \caption{Numbers of bona fide CSOs reported in \citetalias{Paper1} and new ones confirmed in this paper, broken down by how many have spectroscopic redshifts reported.}
    \begin{tabular}{c|c|c|c}
        \hline \hline    
        Literature Source & With z & Total & Percentage with z \\\hline
        \citetalias{Paper1} & 54 & 79 & 68.4\% \\
        This paper & 37 & 65 & 56.9\% \\\hline
    \end{tabular}
    \label{zpercent}
\end{table}

\begin{figure}[h]
    \centering
    \includegraphics[width=0.45\linewidth]{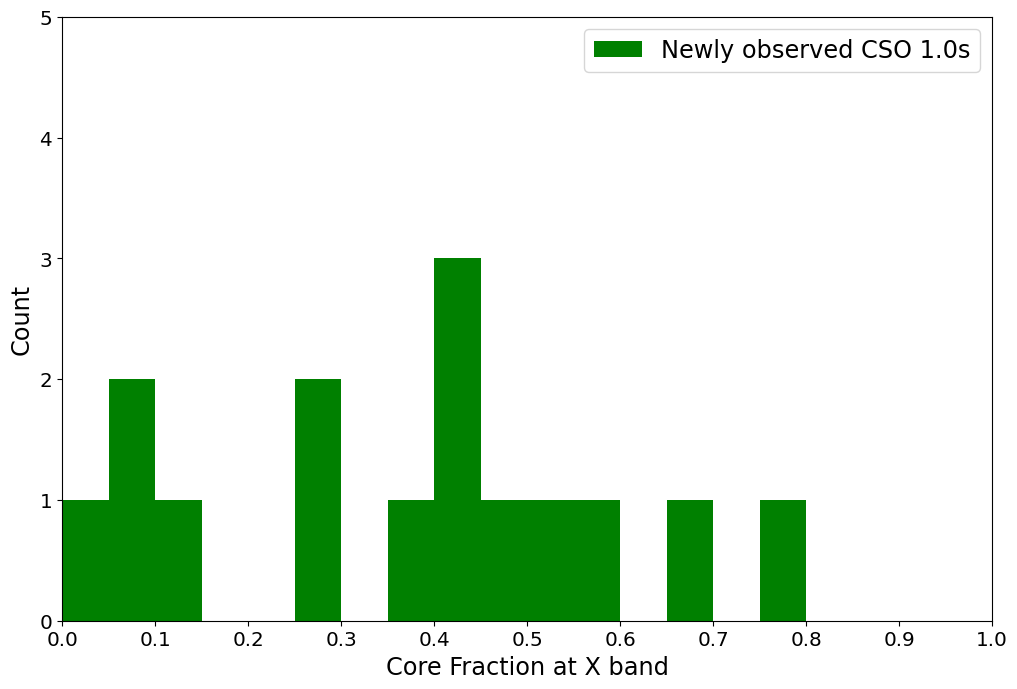} \hfill
    \includegraphics[width=0.45\linewidth]{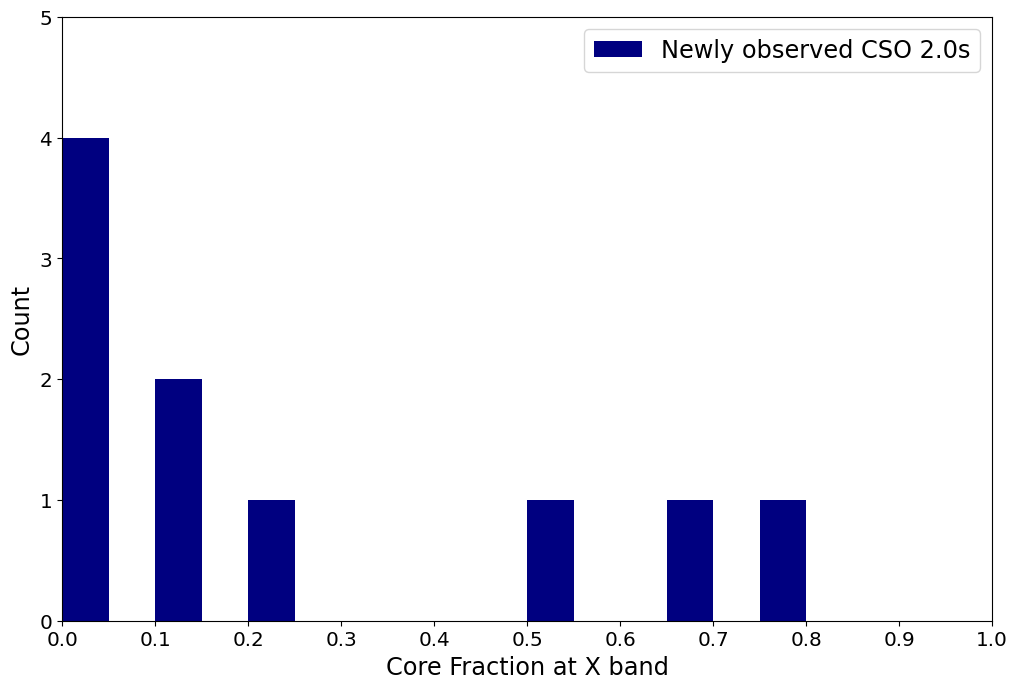} \\
    \includegraphics[width=0.45\linewidth]{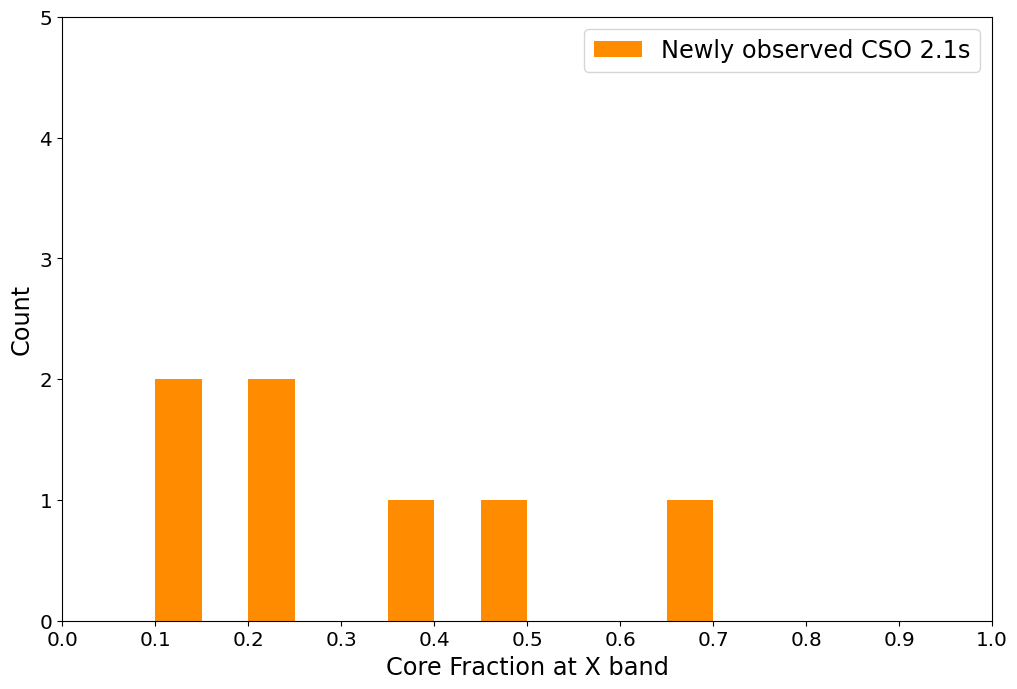} \hfill
    \includegraphics[width=0.45\linewidth]{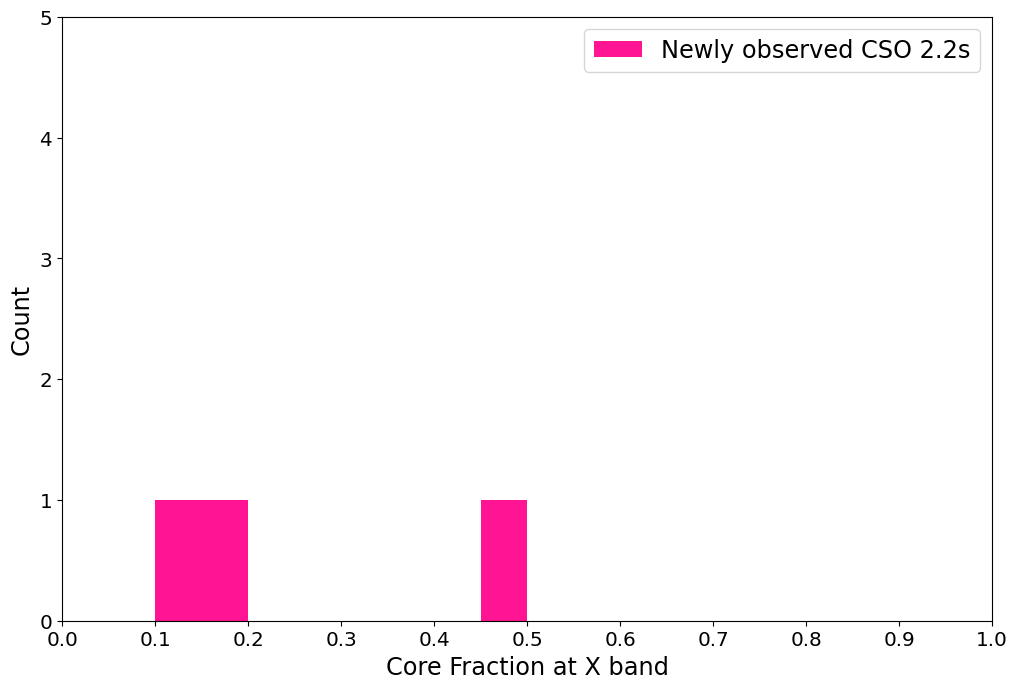}
    \caption{Core fraction distributions of newly observed CSOs sorted by morphological class}
    \label{corefracdists}
\end{figure}

We also plot the redshift of newly observed CSOs against core fraction, but to discuss that graph with the added context of all CSOs, not just those with redshifts, we will discuss the core fraction distributions first, grouped by CSO class (Figure \ref{corefracdists}). We did not measure core fractions for CSOs confirmed in \citetalias{Paper1}, so they do not appear in these distributions. \par
Surprisingly, CSO-1s have a near-uniform distribution in core fraction. We would have expected them to have higher core fractions because they are edge-dimmed. \par
The core fractions of CSO-2s are also a near-uniform distribution, but with a noticeable peak at lower core fractions. A high core fraction can indicate relativistic beaming \citep{Pei_2016} or recently restarted jet activity \citep{Nair_2024}, and since CSO-2s are more likely to have lobe-dominant emission, those with high core fractions immediately stand out. They are J0015-1807 (core fraction = 0.543 $\pm$ 0.010), J1052+8317 (0.671 $\pm$ 0.003), and J1256+5652 (0.759 $\pm$ 0.040). All these sources either have dim lobes or are barely resolved, possibly causing emission from multiple components to overlap, making it difficult to ensure that we are only capturing the core emission. The faintness of the lobes are likely a combination of them being intrinsically dim and our lack of short spacings in our (u,v) coverage, which could lead to us underestimating the largest angular size of some sources. \par 
Of our confirmed CSO-2.1s and 2.2s, only seven and three of them, respectively, have detected cores, so we urge caution in interpreting the results. The highest core fraction CSO-2.1, J1205+2031 (0.692 $\pm$ 0.003), has a large discrepancy between the spatial extents of its lobes. This could imply that the counter jet is being Doppler boosted away and that if the source were more inclined, we would not detect the jet at all, or it could also indicate differences in the environment around each lobe. \par

\begin{figure}[h]
    \centering
    \includegraphics[width=0.45\linewidth]{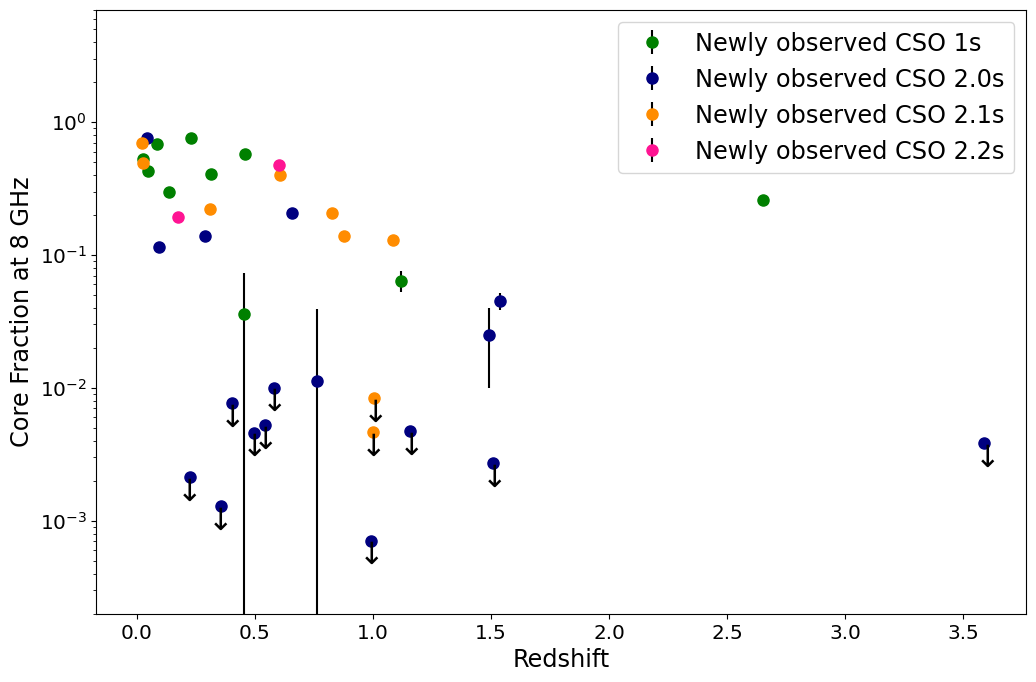}
    \caption{Redshift versus core fraction of newly observed CSOs with identified cores, sorted by CSO class}
    \label{zcorefrac}
\end{figure}

Finally, we look at the redshift distribution of newly observed CSOs by core fraction (Figure \ref{zcorefrac}). We notice an absence of high redshift, high core fraction CSOs, with the exception of J1258+5421, which is a CSO-1 with a high core fraction and high redshift. We should see an observational bias toward sources with higher redshifts having higher core fractions. This is because lobe components are steep spectrum and therefore would become dimmer faster than core components as redshift increases, lowering the overall integrated flux density of the source while keeping the core flux density roughly the same (unless it increases in the case of inverted-spectrum cores). The fact that we don't detect any sources with high redshift and high core fraction except J1258+5421 is an unexpected result, and could indicate that there is an actual dearth of these types of sources. \par
We acknowledge that to truly analyze the physical properties of a source, we would account for the (1+z) factor difference between observed and rest frequencies and adjust the flux densities of its components based on their spectral indices. However, an analysis of this kind would be outside the scope of this paper since our sources generally have complicated morphology and complex spectral indices, and due to the limited bandwidth of our VLBA observations, we cannot place a robust quantitative estimate on how a component's flux density will change as a function of frequency.
\vfill
\vfill
\subsection{Size Distributions} \label{sizedists}
We include only CSOs with spectroscopic redshifts in our linear size plots, including those with linear sizes derived from the literature. 

\begin{figure}[h]
    \centering
    \includegraphics[width=0.45\linewidth]{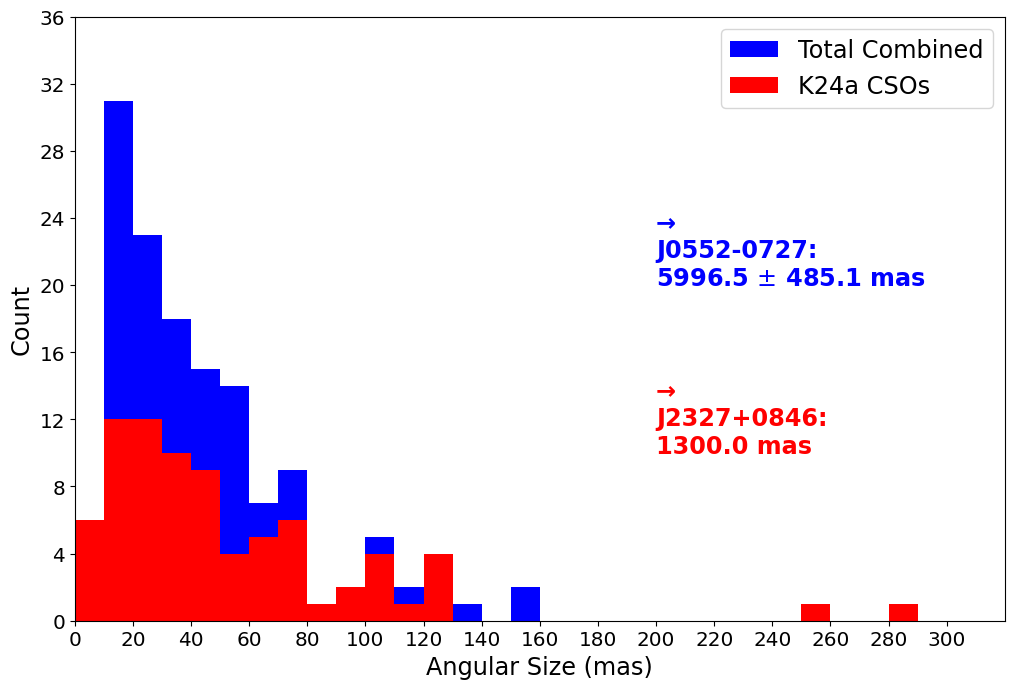} \hfill
    \includegraphics[width=0.45\linewidth]{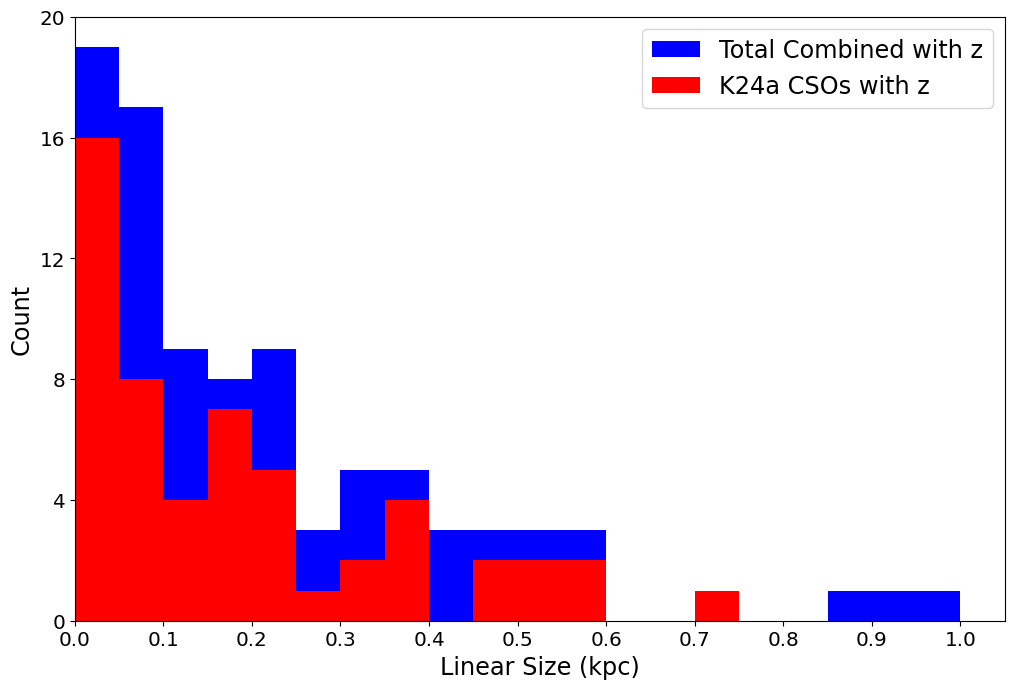}
    \caption{Angular and linear size distributions for CSOs from \citetalias{Paper1} and in total. Only those with reported spectroscopic redshifts are included in the linear size plot. Off the angular size chart are J2327+0846 (\citetalias{Paper1}) and J0552-0727 (this paper).}
    \label{anglinsize}
\end{figure}

The angular and linear size distributions of CSOs from \citetalias{Paper1} and newly confirmed ones are shown in Figure \ref{anglinsize}. Most of the angular sizes measured in this paper are below 80 mas, similar to those in \citetalias{Paper1}. As pointed out in \citetalias{Paper1}, there is a strong selection effect due to the limited sensitivity of our observations to structures larger than 100 mas. This, together with the fact that CSOs with large angular sizes, such as J0552-0727 and J2327+0846, have been identified could indicate there is a population of CSOs that have been missed due to selection effects. This motivates future observing campaigns with better (u,v) coverage at short spacings. We also cannot identify CSOs with an angular size smaller than 10 mas due to angular resolution limitations. \par
Our linear size distribution shows that many of our sources are smaller than 0.1 kpc, though we identify substantially fewer sources smaller than 0.05 kpc than \citetalias{Paper1}. The distribution of our newly confirmed sources appears more uniform at larger sizes before tapering off at about 0.6 kpc. There are clearly selection effects at work here and a definitive discussion of the sizes of the CSOs in our sample must await lower frequency VLBI observations.

We perform two different statistical tests to investigate if the 500 pc size cutoff found in \citetalias{Paper2} is still supported with the addition of the newly confirmed CSOs. These are a one-sample Kolmogorov-Smirnov (KS) test and a binomial test, both comparing all CSO-2s in the 5 GHz Pearson-Readhead complete sample (\citealt{PR}, hereafter \citetalias{PR}), the First Caltech-Jodrell Bank VLBI survey (\citealt{CJ1}, hereafter \citetalias{CJ1}), and the 2.7 GHz Peacock-Wall complete sample (\citealt{PWa,PWb}, hereafter \citetalias{PWa}; \citetalias{PWb}) that have reported spectroscopic redshifts against a uniform distribution of linear sizes. We confirm two more sources included in \citetalias{CJ1} as CSOs: J0650+6001 and J1845+3541, but only J1845+3541 is a CSO-2 (specifically a 2.0). We achieve a p-value of 7.0 $\times$ 10$^{-5}$ for our KS test and 1.4 $\times$ 10$^{-4}$ for our binominal test. Both tests yielded slightly lower p-values than in \citetalias{Paper2}. These high levels of significance strongly suggest we cannot ignore the tendency for CSO-2s to occupy lower linear sizes than predicted by uniform jet expansion speed over their lifetimes, assuming the creation times of CSOs form a uniform distribution.

\subsection{Spectral Peaks in CSOs}
\begin{figure}[h]
    \centering
    \includegraphics[width=0.45\linewidth]{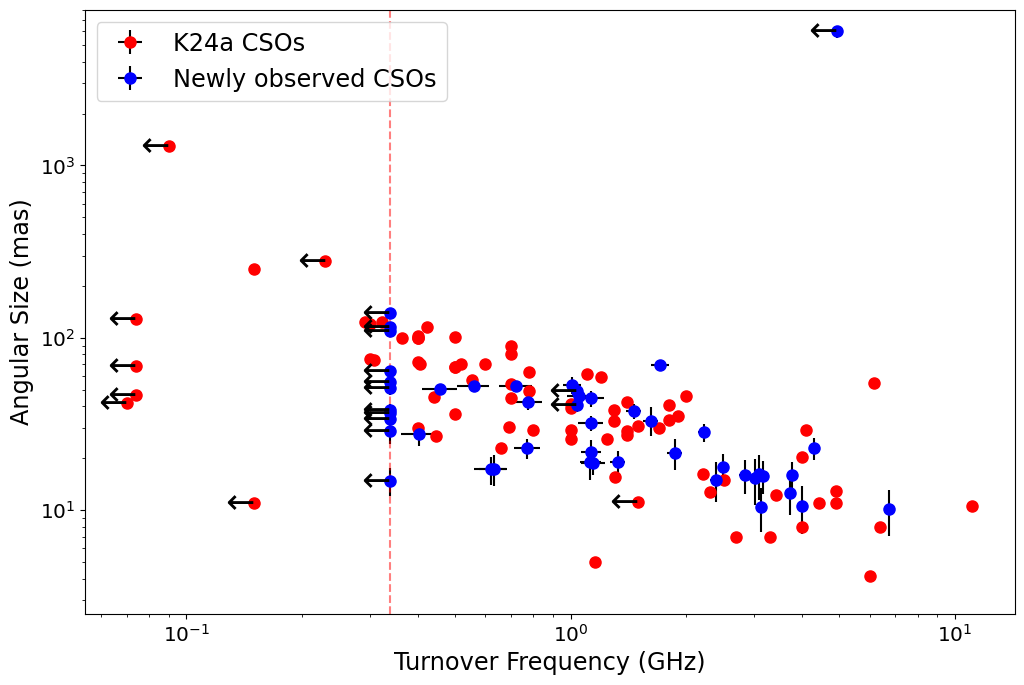} \hfill
    \includegraphics[width=0.45\linewidth]{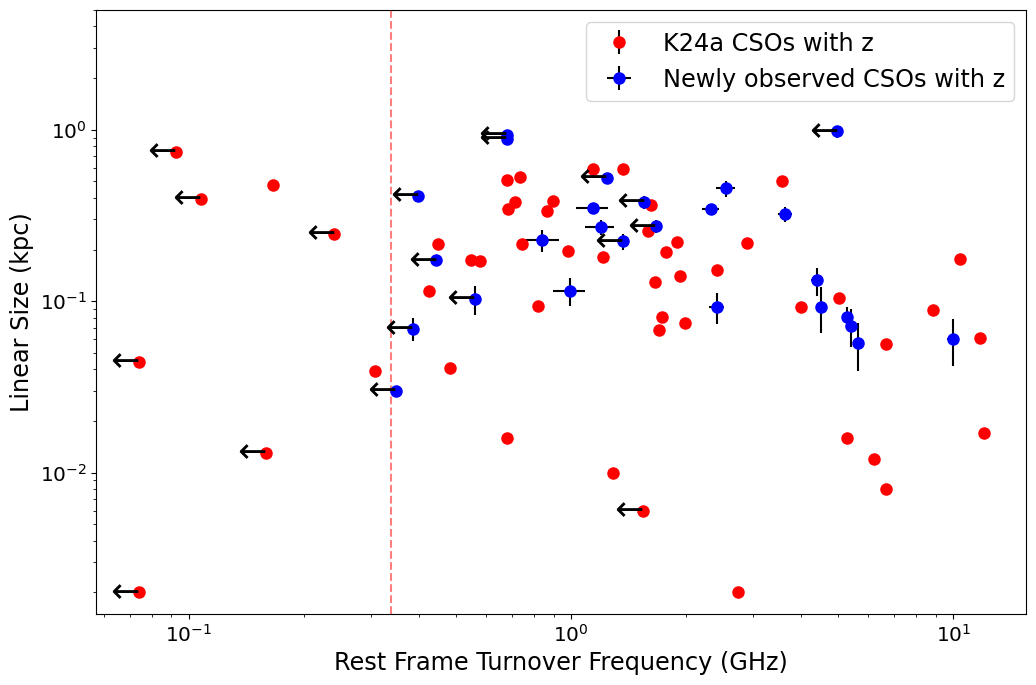}
    \caption{Frequency of spectral peak versus angular and linear size of all confirmed CSOs. The turnover frequency has a (1+z) correction for rest frame frequency applied in the linear size plot. Only those with reported spectroscopic redshifts are included in the linear size plot. Arrows on points indicate upper limits. The lengths of the arrows are not representative of the uncertainty. The dotted red line depicts the lower limit of our VLA data (0.3385 GHz).}
    \label{sizeturnoverfreq}
\end{figure}

\begin{figure}[h]
    \centering
    \includegraphics[width=0.45\linewidth]{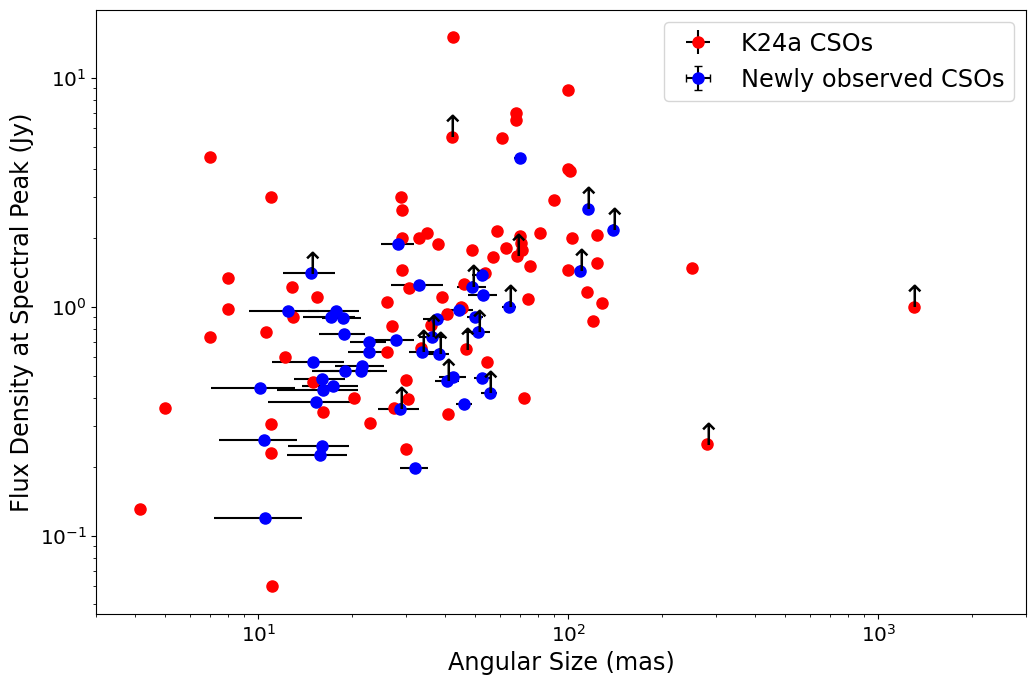} \hfill
    \includegraphics[width=0.45\linewidth]{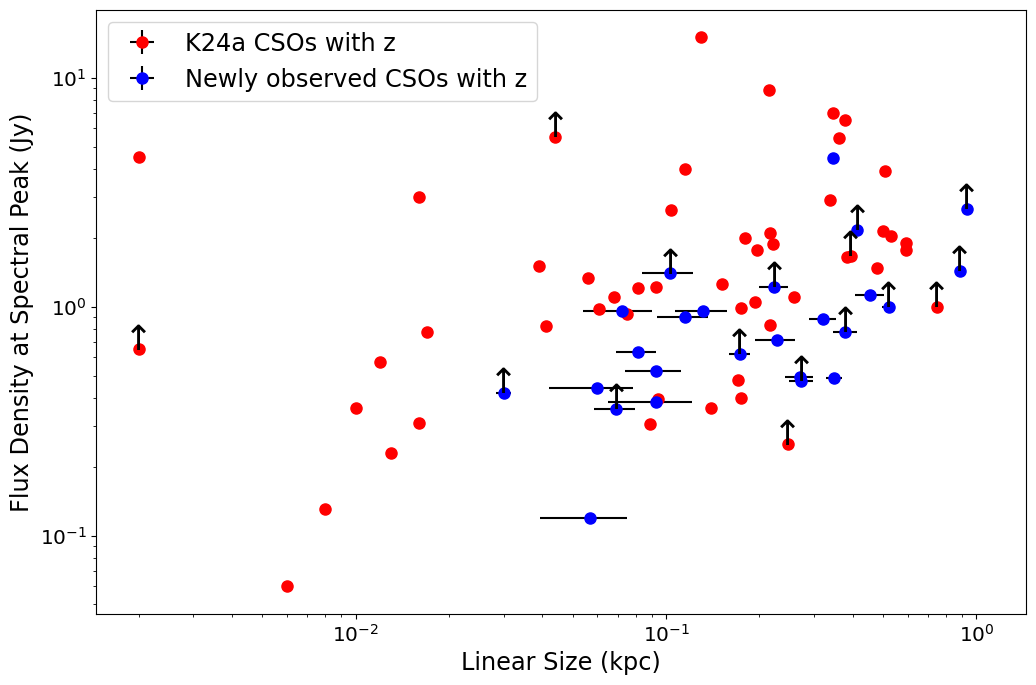}
    \caption{Angular and linear size versus turnover flux density of all confirmed CSOs. Only those with reported spectroscopic redshifts are included in the linear size plot. Arrows on points indicate lower limits. The lengths of the arrows are not representative of the uncertainty.}
    \label{sizeturnoverflux}
\end{figure}

\begin{figure}[h]
    \centering
    \includegraphics[width=0.5\linewidth]{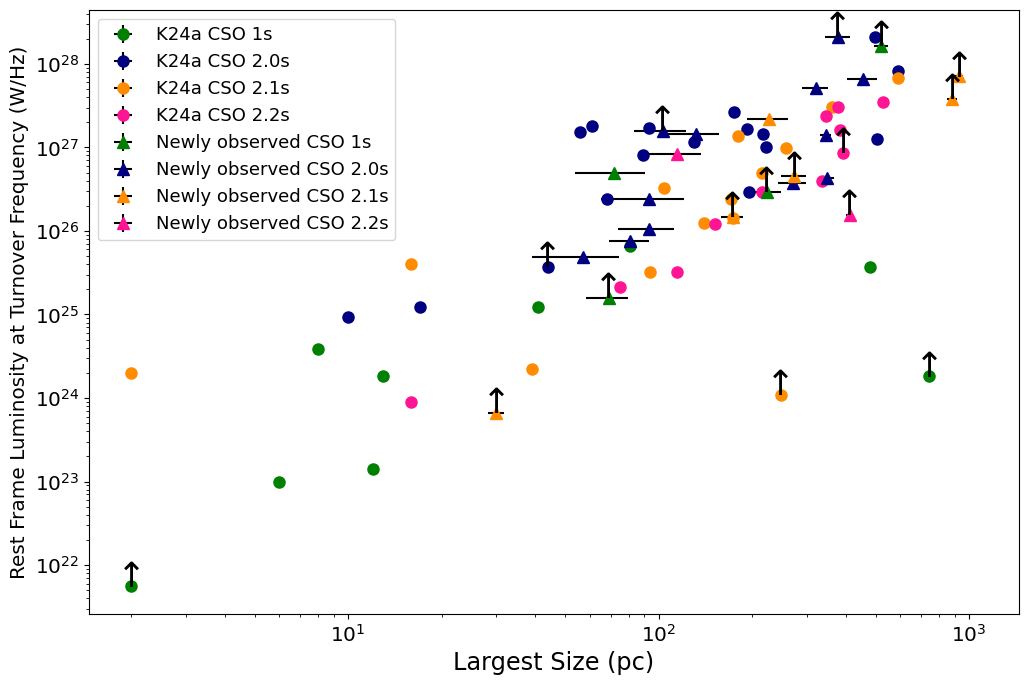}
    \caption{Largest linear size versus rest frame luminosity at turnover frequency of \citetalias{Paper1} CSOs and newly observed CSOs with reported spectroscopic redshifts. Arrows on points indicate upper limits. The lengths of the arrows are not representative of the uncertainty.}
    \label{THEBIGPLOT}
\end{figure}

In Figures \ref{sizeturnoverfreq}, \ref{sizeturnoverflux}, and \ref{THEBIGPLOT}, we look for trends among CSOs based on their spectral peaks and sizes. The angular size plots of Figures \ref{sizeturnoverfreq} and \ref{sizeturnoverflux} of the newly observed bona fide CSOs are in good agreement with those of the bona fide CSOs of \citetalias{Paper1}. On the linear size plots, there are fewer newly observed CSOs at sizes below $\sim$0.03 kpc and at higher flux densities, which is likely due to limited angular resolution and the fact that we are probing a dimmer population of CSOs. 
We were able to measure turnover frequencies above 340 MHz (within 1$\sigma$) or upper limits for 47 of the newly observed CSOs, of which 26 had reported redshifts. Figure \ref{THEBIGPLOT} is an update of Figure 1 from \citetalias{Paper3}, showing the rest frame luminosity at our sources' turnover frequencies compared with their linear sizes. The equation for rest frame luminosity comes from \citet{Hogg_1999} and is 
\begin{equation}
    L_{\nu(1+z)} = 4 \pi {D_A}^2 S_\nu (1+z)^3.
\end{equation}

The new CSO-2.0s and CSO-2.2s generally occupy the same regions, though two CSO-2.0s, J1442+3042 (270 $\pm$ 28 pc) and J1700+3830 (348 $\pm$ 20 pc), approach the region populated by 2.2s. Three of the new CSO-1s (J0242-2132 (223 $\pm$ 24 pc), J0650+6001 (72 $\pm$ 18 pc), J1258+5421 (521 $\pm$ 26 pc)) extend further upward than before. The most luminous of these, J0650+6001 and J1258+5421, share the region of the graph occupied mostly by CSO-2.0s. J0650+6001 has one of the lowest core fractions of any of our CSO-1s and has fewer resolved components than some other sources. It is possible, then, that future observations may find that J0650+6001 is actually a CSO-2.0 that we are not able to fully resolve with the VLBA in this work. J1258+5421, on the other hand, is a well-resolved source with an average core fraction for a CSO-1. This may again be evidence that CSO-1s are a more diverse set of sources than originally thought. \par

\vfill
\subsection{VLBA \& VLA Flux Density Comparison}

\begin{figure}[h]
    \centering
    \includegraphics[width=0.45\linewidth]{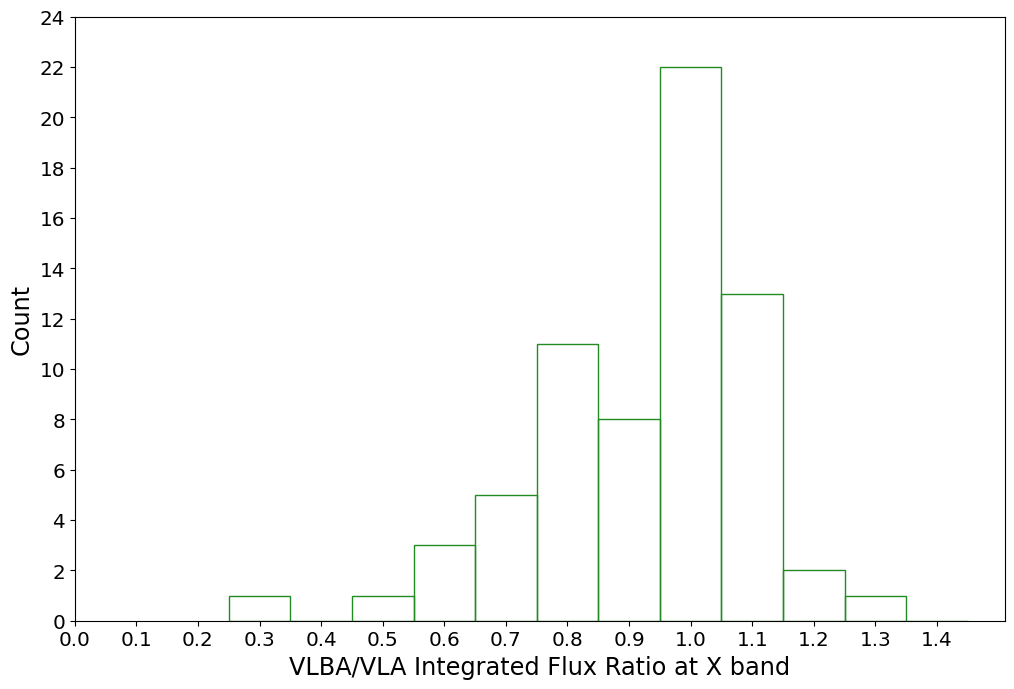} \hfill
    \caption{Distribution of ratios of total integrated flux densities of CSOs observed with the VLBA and VLA. For J0552-0727 (Ratio = 0.793), since we only resolved the core in the VLBA observations, only the integrated flux densities of the core emission are compared. J1326+5712 is excluded since it was not included in the VLA observations.}
    \label{vla_vlba_fluxcomp}
\end{figure}

Barring variability, we expect a source to have an equal or higher flux density on the VLA than on the VLBA. Thus, a comparison of the total flux densities measured on the VLA and on the VLBA is a good way to assess whether there is structure that has been completely resolved out on the VLBA. We compare the integrated flux densities of our VLBA and VLA images at 8 GHz using the VLA images we made in the 8.11225-8.6242 GHz band, with those of our VLBA observations in this band. The only source that we treat differently is J0552-0727 since we can only resolve its core with the VLBA. Therefore, we only compare the integrated flux density of its core in the VLA image to its total integrated flux density from the VLBA image. See Figure \ref{vla_vlba_fluxcomp} for the results of the comparison. \par
We see from Figure \ref{vla_vlba_fluxcomp} that most of the ratios are gathered around 1, with a tail petering downward toward low ratios and a sharp cutoff at higher ones. The outlier CSOs with the lowest flux density ratios are J1215+1730 (Ratio = 0.478 $\pm$ 0.013) and J1755+6236 (0.281 $\pm$ 0.007). Unsurprisingly, these are both sources with large-scale structure in their lobes that is being resolved out. Sources with ratios above 1 are likely due to calibration offsets.

\vfill
\subsection{VLBA Observational Limitations}
\begin{figure}[h]
    \centering
    \includegraphics[width=0.45\linewidth]{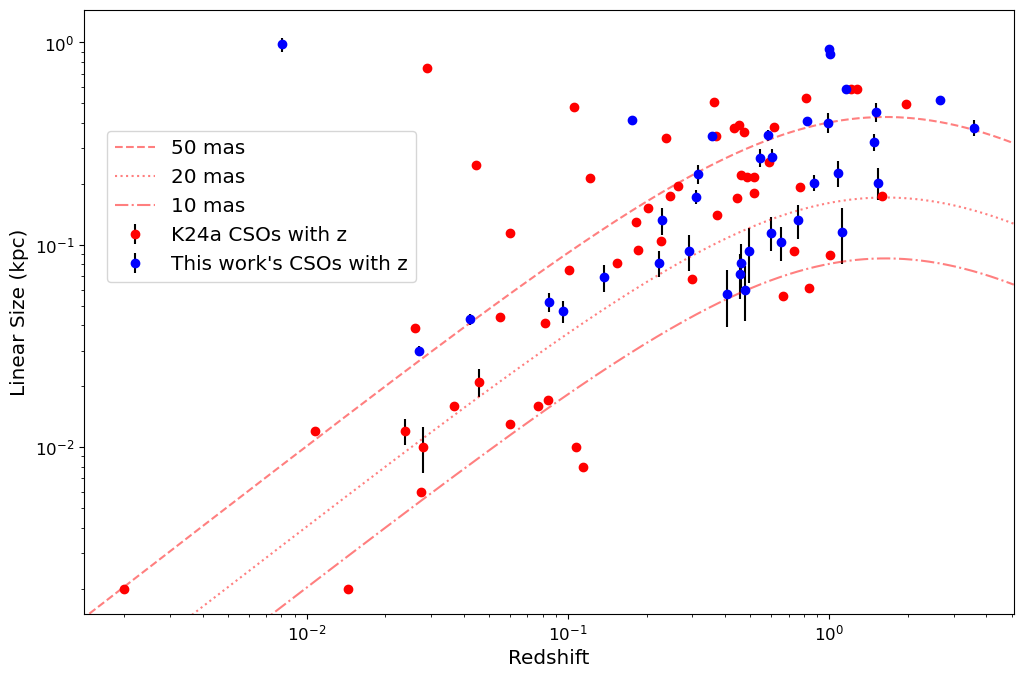}
    \caption{Redshift versus linear size of all confirmed CSOs. Lines showing 50, 20, and 10 mas are added.}
    \label{zlin}
\end{figure}

It is important to emphasize the observational biases that apply to this work due to the lower and upper limits on the baseline lengths of the VLBA and the fact that the lowest observed frequency was 5 GHz. The effects are illustrated in Figure \ref{zlin}, which shows two relatively empty regions: in the top left, and in the bottom right. We are mostly limited by the VLBA brightness sensitivity in the top left region. This region is populated by CSOs from \citetalias{Paper1} imaged by other telescopes with fuller (u,v) coverage, like ``VLBA-Plus" arrays\footnote{https://science.nrao.edu/facilities/vlba/docs/manuals/oss2024A/vlba-plus}, the VLA, the European VLBI Network (EVN), and the Multi Element Remotely Linked Interferometer Network (MERLIN). In fact, the blue point in the very top left is the newly confirmed CSO we required VLA images of to observe -- J0552-0727. The absence of points at the bottom right is likely due to the limits in angular resolution of the VLBA, though it is worth noting that the VLBA provides some of the highest angular resolution available. We classified many sources as indeterminate because we could not adequately detect a spectral index distribution or morphology consistent with a particular CSO class beyond reasonable doubt. Some examples are J0756+6347 and J2244+2600. A few bona fide CSOs have a small enough angular size that they are barely resolved, such as J0428+3259.

\section{Wide-Angle Tail CSOs} \label{htcsos}
\begin{figure}[h]
    \centering
    \includegraphics[page=87,trim={4.35in 6.75in 2.05in 1.85in},clip,width=0.45\linewidth]{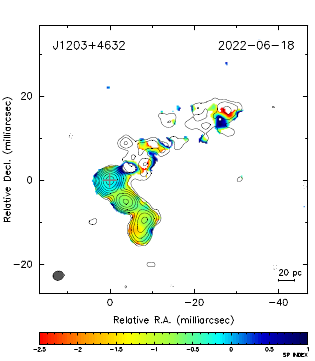} \hfill
    \includegraphics[page=113,trim={4.35in 6.75in 2.05in 1.85in},clip,width=0.45\linewidth]{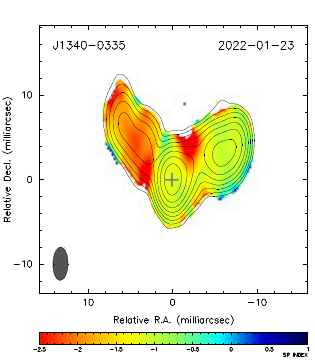} \\
    \includegraphics[page=134,trim={4.35in 6.75in 2.05in 1.85in},clip,width=0.45\linewidth]{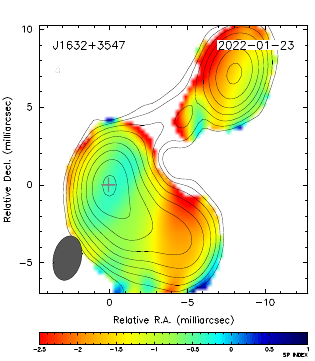} \hfill
    \includegraphics[page=143,trim={4.35in 6.75in 2.05in 1.85in},clip,width=0.45\linewidth]{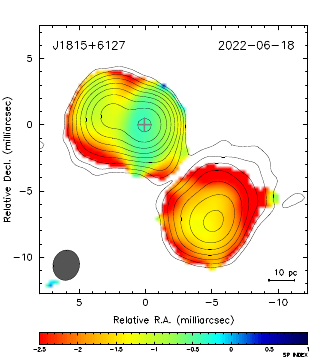}
    \caption{CSOs verified this paper that exhibit a WAT-like morphology. All spectral index maps cover 5-8 GHz and have 5 GHz contours overlaid.}
    \label{wats}
\end{figure}

Four of our CSOs (J1203+4632, J1340-0335, J1632+3547, and J1815+6127) exhibit morphology similar to a WAT (Figure \ref{wats}), meaning their jets appear angled $>90^{\circ}$ from each other but still in a common direction \citep{Mao_2010}. Notably, this is a characteristic that spans CSO classes, as J1340-0335 and J1632+3547 are CSO-1s, J1203+4632 is a CSO-2.1, and J1815+6127 is a CSO-2.2. These are also CSOs that do not exhibit typical S-symmetry or double symmetry, perhaps suggesting a jet perturbation through environmental interactions. Individual source descriptions and reasons for this categorization are given in Appendix \ref{indivsources}. The characteristics and importance of head-tail (HT) radio sources are well described in a series of articles based on their radio structure \citep{1975A&A....38..381M,1977Natur.265..315V,1978ApJ...226L.119O,1979ApJ...229L..59O,1981ApJ...246L..69R,1985A&A...143..136B,1985AJ.....90..954O,1985ApJ...295...80O,1986ApJ...307...73B,1986ApJ...301..841O,1987ApJ...316..113O,1989A&AS...79..391D,1989MNRAS.240..501L}.
\par
HT galaxies often dominate their clusters \citep{Owen_1976} and are much larger than 1 kpc, with emission on arcsecond scales (e.g., \citealt{Pal_2023}). Therefore, it is of interest to find so many CSOs with WAT-like emission. Multi-frequency follow-up in the infrared, optical, and X-ray wavelengths would be most interesting, and might reveal evidence of relative motions between the host galaxy and local environment. \par

The simple model of HT sources (e.g., \citealt{1979Natur.279..770B}) posits that the source of the two jets lies in the nucleus of a cluster galaxy moving through the intracluster medium. Once the jets escape the galaxy, the ram pressure is sufficient to deflect them so that they become trails moving slowly through this medium, emitting until the relativistic electrons cool. Clearly, this is not a viable explanation for WAT-like CSOs that are only $\sim$1 kpc in size. However, a natural extension of this idea is that the deflection is due to the motion of a minor galaxy with an active, jet-forming nucleus through the interstellar medium of a major galaxy with which it is merging. 

Typically, merging galaxies with masses $\gtrsim10^6\,{\rm M}_\odot$ get progressively stripped as they spiral together under dynamical friction, creating gas and stellar streams. A dense core fueling a jet-forming nucleus in a minor galaxy can survive for hundreds of million years until it merges with the nucleus of the major galaxy, which may or may not have an active nucleus. It is most plausibly a spiral where the medium should be denser than in an elliptical galaxy. A simple estimate of the orbital speed is $\sim300\,{\rm km\,s}^{-1}$ and the density is $\sim10^{-24}\,{\rm g\,cm}^{-3}$, so the ram pressure is $\sim10^{-9}\,{\rm dyne\,cm}^{-2}$. This is sufficient to deflect a CSO-1 to give it WAT-like jet trails.

A general consequence of this interpretation is that these sources should be offset from the centers of their host optical galaxies by perhaps ten times their sizes. Also, they need not be in clusters. This explanation is therefore quite refutable. If it is substantiated, then more detailed multi-spectral investigations will be highly instructive. If it is not, then a WAT-like CSO morphology presumably reflects a complex gas flow within a single host galaxy.

\vfill
\section{MSOs} \label{msos}
\begin{figure}[h]
    \centering
    \includegraphics[page=68,trim={4.35in 6.75in 2.05in 1.85in},clip,width=0.45\linewidth]{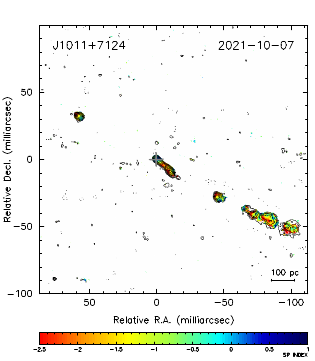}
    \caption{5-8 GHz spectral index map of J1011+7124, newly confirmed as an MSO in this paper.}
    \label{j1011_mso}
\end{figure}

\begin{figure}[h]
    \centering
    \includegraphics[page=75,trim={4.35in 6.75in 2.05in 1.85in},clip,width=0.45\linewidth]{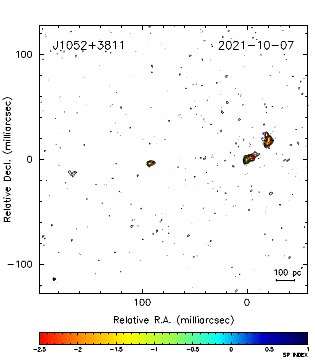} \hfill
    \includegraphics[page=164,trim={4.35in 6.75in 2.05in 1.85in},clip,width=0.45\linewidth]{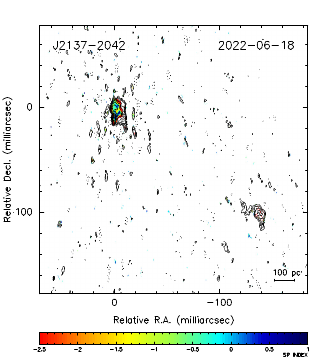}
    \caption{5-8 GHz spectral index maps of J1052+3811 and J2137-2042, which are both indeterminate MSO candidates.}
    \label{msos_indet}
\end{figure}

Medium symmetric objects, or MSOs, are another subclass of radio AGN. They are classified identically to CSOs except they are $>$1 kpc but smaller than traditional double galaxies (\citealt{Fanti_1995}, \citetalias{1996ApJ...460..612R}). Out of our three sources rejected by size (J1011+7124, J1052+3811, and J2137-2042) only J1011+7124 has an obvious flat-spectrum core component with symmetrical steep-spectrum lobe emission (Figure \ref{j1011_mso}). J1052+3811 does not have an obvious core location and is therefore an indeterminate MSO candidate (Figure \ref{msos_indet}). J2137-2042 lies at a negative declination, meaning it was observed with relatively poor (u,v) coverage. It exhibits two diffuse components that could be lobes, but since our 5 GHz map of it has a high RMS noise that is distorting the spectral index map, we can't be sure of the component at the phase center and therefore classify this source as an indeterminate MSO candidate (Figure \ref{msos_indet}). \par
\citetalias{Paper2} used the relative populations of CSO-2s, MSOs/CSSs, and Fanaroff-Riley galaxies \citep{Fanaroff_1974}, i.e., FRIs and FRIIs, in complete flux density-limited samples as evidence that most CSO-2s do not evolve into larger radio galaxies. The positive identification of an MSO will contribute to similar statistical analyses carried out in the future comparing populations of radio AGN.

\vfill
\section{Supermassive Binary Black Hole Candidates} \label{sbbhcands}
\begin{figure}[h]
    \centering
    \includegraphics[page=39,trim={1.95in 4.2in 4.45in 4.6in},clip,width=0.45\linewidth]{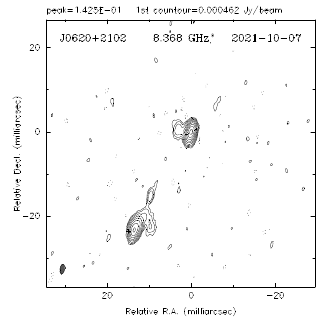}
    \caption{8 GHz Stokes I contour map of J0620+2102. We do \textit{not} consider this an SBBH candidate despite its extended emission because its lobes can be logically traced back to an undetected core in between them.}
    \label{j0620example}
\end{figure}

As stated in Section \ref{intro}, identifying SBBHs can help us answer many physics questions surrounding them, and compact AGN such as CSOs are ideal targets to study them with. There are two signatures of an SBBH: (i) two compact flat/inverted-spectrum components, and (ii) lobe emission that cannot be traced back to one center of emission, indicating multiple jets. Detecting either of these qualities can indicate an SBBH candidate, but a more definitive classification would require kinematic measurements of the suspected core components that identify them as orbiting each other. Note that for criterion (ii), SBBH lobe emission is distinct from simple ``extended emission" that is seen in many CSO-2s and is a defining characteristic of a CSO-2.2. For example, although J0620+2102 has extended lobe emission from its northern lobe detected at three contour levels, this can be easily explained by environmental interactions, and its two lobes can be traced back to an area directly between them (Figure \ref{j0620example}). \par

\begin{figure}[h]
    \centering
    \includegraphics[page=13,trim={4.35in 6.75in 2.05in 1.85in},clip,width=0.45\linewidth]{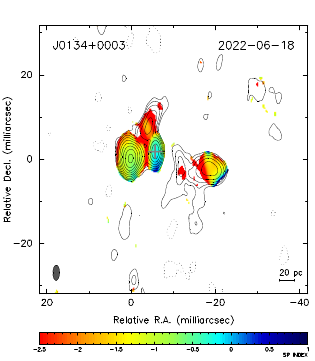} \hfill
    \includegraphics[page=87,trim={4.35in 6.75in 2.05in 1.85in},clip,width=0.45\linewidth]{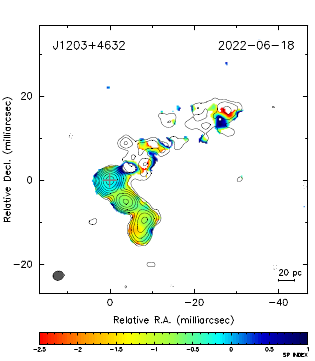} \\
    \includegraphics[page=135,trim={4.35in 6.75in 2.05in 1.85in},clip,width=0.45\linewidth]{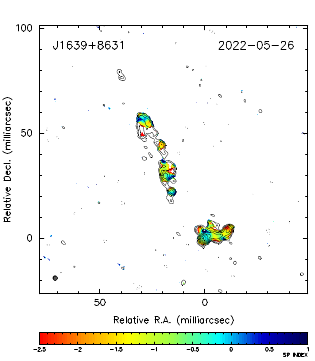} \hfill
    \includegraphics[page=172,trim={4.35in 6.75in 2.05in 1.85in},clip,width=0.45\linewidth]{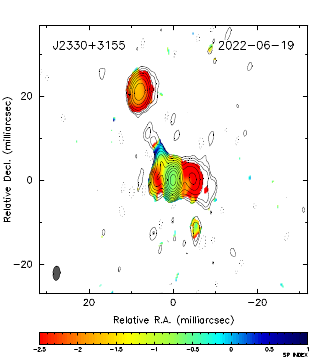}
    \caption{5-8 GHz spectral index maps of newly confirmed CSOs that may be SBBH candidates. 5 GHz contours are overlaid and start at three times the RMS noise level. A grey cross has been placed on what we believe is the core component if we detect it.}
    \label{sbbhcandsfig}
\end{figure}

We examined all of our sources using our SBBH candidate criteria and found four newly identified bona fide CSOs worth further investigation due to prominent off-axis emission: J0134+0003, J1203+4632 (also a WAT-like CSO), J1639+8631, and J2330+3155 (see Figure \ref{sbbhcandsfig}). Unfortunately, none of these sources are strong SBBH candidates since none of them exhibit more than one compact flat/inverted spectrum core. In Appendix \ref{indivsources}, we discuss each of these sources in more detail. Further observations to obtain kinematic data are needed to investigate the possible SBBH nature of these objects. Follow-up would be useful not only for SBBH verification, but also for CSO verification using our v$_{app} <$ 2.5 c criterion and providing more insight into early-life AGN radio jets.

\vfill
\section{High Lobe Flux Ratio CSOs} \label{highfluxratio}
\begin{figure}[h]
    \centering
    \includegraphics[page=56,trim={4.35in 4.04in 2.05in 4.56in},clip,width=0.45\linewidth]{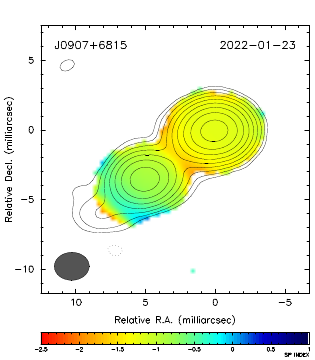} \hfill
    \includegraphics[page=87,trim={4.35in 6.75in 2.05in 1.85in},clip,width=0.45\linewidth]{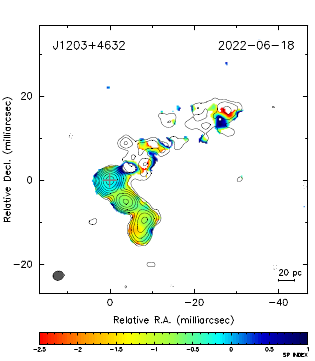} \\
    \includegraphics[page=126,trim={4.35in 6.75in 2.05in 1.85in},clip,width=0.45\linewidth]{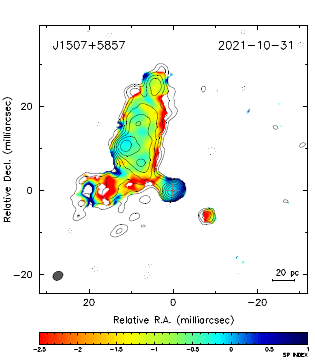} \hfill
    \includegraphics[page=167,trim={4.35in 6.75in 2.05in 1.85in},clip,width=0.45\linewidth]{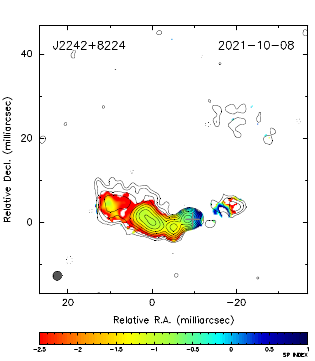}
    \caption{Sources with high lobe flux ratios. Shown are a 5-15 GHz map of J0907+6815 and 5-8 GHz spectral index maps of J1203+4632, J1507+5857, and J2242+8224. 5 GHz contours are overlaid on all maps at three times the RMS noise level. Grey crosses indicate where we suspect the core is if we detect it.}
    \label{fluxratiosources}
\end{figure}

Four of our sources test the boundaries of the ``symmetric" part of ``CSO" due to the difference in the flux densities of the components straddling the nucleus, which could indicate relativistic beaming: J0907+6815, J1203+4632, J1507+5857, and J2242+8224. See Figure \ref{fluxratiosources} for spectral index maps of all four. To estimate the flux density ratios of the lobes of these sources, we used the same Gaussian components that we fit when finding each source's core fraction, took the sums of the integrated flux densities of components making up each of the two lobes, then divided the brighter one by the dimmer one. J0907+6815 does not have a definitively identified core, so we estimated its lobe flux ratio with the core at the center of the southeastern component (see Section \ref{j0907_indiv} for more details). See Table \ref{fluxratios} for our estimates. \par

\begin{table}[h]
    \centering
    \caption{Estimates of the integrated flux ratios of the lobes of the four sources discussed in this section.}
    \begin{tabular}{c|c}
        \hline \hline
        Source Name & Flux Ratio at 8 GHz \\\hline
        J0907+6815\tablenotemark{a} & 22.4 \\
        J1203+4632 & 6.49 \\
        J1507+5857 & 91.8 \\
        J2242+8224 & 79.9\\\hline
    \end{tabular}
    \tablenotetext{a}{No definitive core identification}
    \tablecomments{All measurements are based off of 8 GHz maps}
    \label{fluxratios}
\end{table}

Another possible indicator of relativistic beaming is a high core fraction \citep{Pei_2016}. However, a high core fraction is not necessarily strongly correlated with relativistic beaming and there are clearly other affecting factors. For example, if a jet from an edge-on radio galaxy is strongly bent toward the observer through some environmental interaction, the jet's emission could appear beamed without affecting the core fraction. Our estimated core fractions for these high lobe flux ratio sources are: 0.399 $\pm$ 0.006 for J1203+4632, 0.221 $\pm$ 0.008 for J1507+5857, and 0.0637 $\pm$ 0.0058 for J2242+8224. We are not able to pinpoint the core location in J0907+6815, so its upper limit is 3.68 $\times$ 10$^{-3}$. Of these four, J1203+4632 and J1507+5857 have by far the highest core fractions. \par

A sample of blazars studied by \citet{Pei_2016} had core fractions spanning a range of 6.9178 $\times$ 10$^{-5}$ to 0.99985 (log$R$ from -4.16 to 3.83, where $R = S_{core}/S_{ext}$, the ratio of the core to extended emission), and an average of 0.324$^{+0.567}_{-0.296}$ (log$R$ = -0.32 $\pm$ 1.23). Since this is such a wide range, it is hard to say anything definitive about how beamed any of our sources are based on their core fractions.\par

These sources should be monitored since core-dominated CSOs are more likely to be variable \citepalias{Paper1}. For the time being, we retain the CSO classification until further observations prove otherwise.

\vfill
\section{Summary} \label{summary}
We have presented the results of a joint-VLBA and VLA analysis of 167 CSO candidates. This is an extension of the effort to identify new CSOs using spectral index maps from VLBA data, started in \citet{Tremblay_2016} and \citetalias{Paper1}. Among the A-candidate CSOs selected in \citetalias{Paper1}, we confirm 65 of them as bona fide CSOs, a confirmation rate of about 39\%. Adding these to the ones verified in \citetalias{Paper1} equates to a total sample size of 144 bona fide CSOs, an increase of about 82\%. By taking supplementary VLA data, we determined radio spectral peak information of our bona fide CSOs and detect a low redshift CSO/MSO (J0552-0727) that we would have otherwise resolved out due to a lack of shorter interferometric spacings on the VLBA. \par
The CSOs observed in this study fit into the population studied in \citetalias{Paper1}, \citetalias{Paper2}, and \citetalias{Paper3}. We replicate many of those results, such as CSO-2s generally being more luminous than CSO-1s, CSO-2.0s occupying a broader range of redshifts that extends higher than that of CSO-2.2s, and CSO-1s having mostly lower redshifts. These trends may be largely due to selection effects because CSO-1s and 2.2s have lower surface brightnesses on average than 2.0s. That being said, there is evidence that CSO-1s represent a more diverse sample of sources than originally found. Since they are generally less luminous than CSO-2s, they were only detected at redshifts less than about 0.2 in \citetalias{Paper1}, but we have now identified CSO-1s up to z = 2.65277. Their core fractions also show a more uniform distribution than would be expected from a population of edge-dimmed sources. In fact, more than half of the newly confirmed CSO-1s have core fractions below 50\%, and several have undetected cores. There are also a few with peak luminosities closer to that of CSO-2.0s. Some of these sources may be reclassified upon the advent of new, higher resolution data that can better resolve their structures, but we anticipate the general trends observed will remain. \par
Some CSOs exhibit unique morphology, including four potential SBBHs, four WAT-like objects, and four objects with radio lobes having very different integrated flux densities. Kinematic studies and multi-messenger observations are necessary for progress with these objects. \par
In this study, we explore a population of dimmer, harder to observe CSOs, and with that comes realizations of limitations in angular resolution and brightness sensitivity of the VLBA. 46 sources, or about 28\%, were given an indeterminate classification given the available data. In most cases, this was a combination of low brightness, small angular extent, limited spectral index maps, and limited (u,v) coverage. We will likely need to employ more innovative observing strategies in the future, such as more on-source time, supplementing observations with interferometers with shorter antenna spacings and better north-south coverage, and phase referencing to study fainter sources. Our sample of 65 bona fide CSOs is biased toward sources that don't require phase referencing to detect. As people continue to observe CSO candidates, we expect the need for observations that include phase referencing will increase. Though these sorts of schedules will not contribute as many new bona fide CSOs to the sample, they will crucially unveil sources in the low-brightness, high redshift regime. We expect that the upcoming Next Generation Very Large Array (ngVLA, \citealt{ngVLA}) will completely revolutionize the study of CSOs due to its unparalleled combination of (u,v) coverage, angular resolution, higher brightness sensitivity, and wider frequency bands. It should not only detect dim, high redshift sources, but also sources with very extended lobe emission or small angular sizes. \par
CSOs are underexplored yet valuable sources that will likely offer key insight into the formation of radio jets in AGN. Now that there are well over 100 identified CSOs, we believe they represent a significant subclass of AGN worth investigating as a priority, rather than a rare phenomenon.

\begin{acknowledgments}
We thank the anonymous reviewer for their many helpful suggestions and insightful comments. \par
This research has made use of the NASA/IPAC Extragalactic Database (NED), which is funded by the National Aeronautics and Space Administration and operated by the California Institute of Technology. \par
This research made use of pandas \citep{McKinney_2010, McKinney_2011}. \par
This research made use of SciPy \citep{Virtanen_2020}. \par
This research made use of matplotlib, a Python library for publication quality graphics \citep{Hunter:2007}. \par
This research made use of NumPy \citep{harris2020array}. \par
Support for this work was provided by the National Science Foundation through award NSF/AST-1835400 and through the Grote Reber Fellowship Program administered by Associated Universities, Inc./National Radio Astronomy Observatory. The National Radio Astronomy Observatory and Green Bank Observatory are facilities of the U.S. National Science Foundation operated under cooperative agreement by Associated Universities, Inc. \par
W.M.P. acknowledges that basic research in radio astronomy at the U.S. Naval Research Laboratory (NRL) is supported by 6.1 Base funding. Construction and installation of VLITE was supported by the NRL Sustainment Restoration and Maintenance fund. \par
S.K. was funded by the European Union ERC-2022-STG - BOOTES - 101076343. Views and opinions expressed are however those of the author(s) only and do not necessarily reflect those of the European Union or the European Research Council Executive Agency. Neither the European Union nor the granting authority can be held responsible for them.
A.S. acknowledges support from the NASA contract to the Chandra X-ray Center NAS8-03060.
\end{acknowledgments}
\vspace{2cm}
\bibliography{references}{}
\bibliographystyle{aasjournal}

\appendix
\setcounter{table}{0}
\renewcommand{\thetable}{\Alph{section}\arabic{table}}
\section{Individual Sources} \label{indivsources}
Here, we detail a few bona fide CSOs that have particular characteristics of interest. This largely consists of CSOs explicitly mentioned in main body of the text. The complete figure set of all 171 sources is available in the online journal.
\subsection{J0015-1807}
We suspect this source's core is at the phase center of the map surrounded on either side by very dim lobes that only appear above the third contour level at 5 GHz and are completely undetected at 15 GHz. Its lobes are edge-brightened, making it a CSO-2.0. It has a high core fraction, especially for a 2.0. Its core does not have as flat/inverted of a spectrum as other core identifications in this study, but it is the brightest compact component that persists across frequency, making it the most likely candidate for the core. 
\subsection{J0134+0003}
We identify this source's core as the flat-spectrum component to the immediate west of the phase center. It has interesting lobe morphology, boasting off-axis emission to the northeast of the core that is detected at up to five contour levels in the 5 GHz map and four levels in the 8 GHz map. This makes us hesitant to dismiss it as an imaging artifact. If it were simply an IGM interaction, we would not expect to see it so close to the core, though this could just be a projection effect. Because it's hard to trace this off-axis emission back to the core cleanly, we weakly consider it an SBBH candidate. Other than that, it is a CSO-2.1 because the lobe without the off-axis emission is more compact.
\subsection{J0205+7522}
This source is an archetypical example of a CSO-2.0; it exhibits symmetrical steep spectrum lobe emission about an undetected core. The fact that the lobes have very closely spaced contours at their edges supports the theory that their center of emission is between them.
\subsection{J0242-2132}
This source features a core at the phase center with flat spectrum emission and edge-dimmed lobes with complex spectral indices. Because its lobes are edge-dimmed, it is a CSO-1. Its peak luminosity is also higher than other CSO-1s given its linear size (Figure \ref{THEBIGPLOT}).
\subsection{J0428+3259}
This is one of our smaller sources, only reaching 60 pc in size at 5 GHz and about 10 mas. As a result, it is very hard to resolve all of its components. We have classified many other sources as indeterminate if they are too small to resolve sufficiently and/or if they have very little variation in spectral index across their spatial extent, making it nearly impossible to distinguish core and lobe components. This source demonstrates the importance of having 15 GHz observations. With them, we are able to identify the compact component at the phase center (easily picked out in 15 GHz map) as the core and extended steep spectrum emission on either side. Because this source is so few resolution elements across, we refrain from giving it a CSO classification (i.e., CSO-1, CSO-2.0, CSO-2.1, or CSO-2.2). It has a $>$50\% core fraction, which could partially be due to the difficulty in resolving individual components, stressing our need for higher resolution observations.

\subsection{J0552-0727} \label{j0552indiv}
\begin{figure}[h]
    \centering
    \includegraphics[page=37,trim={5.1cm 2.8cm 5.1cm 4cm},clip,width=0.68\linewidth]{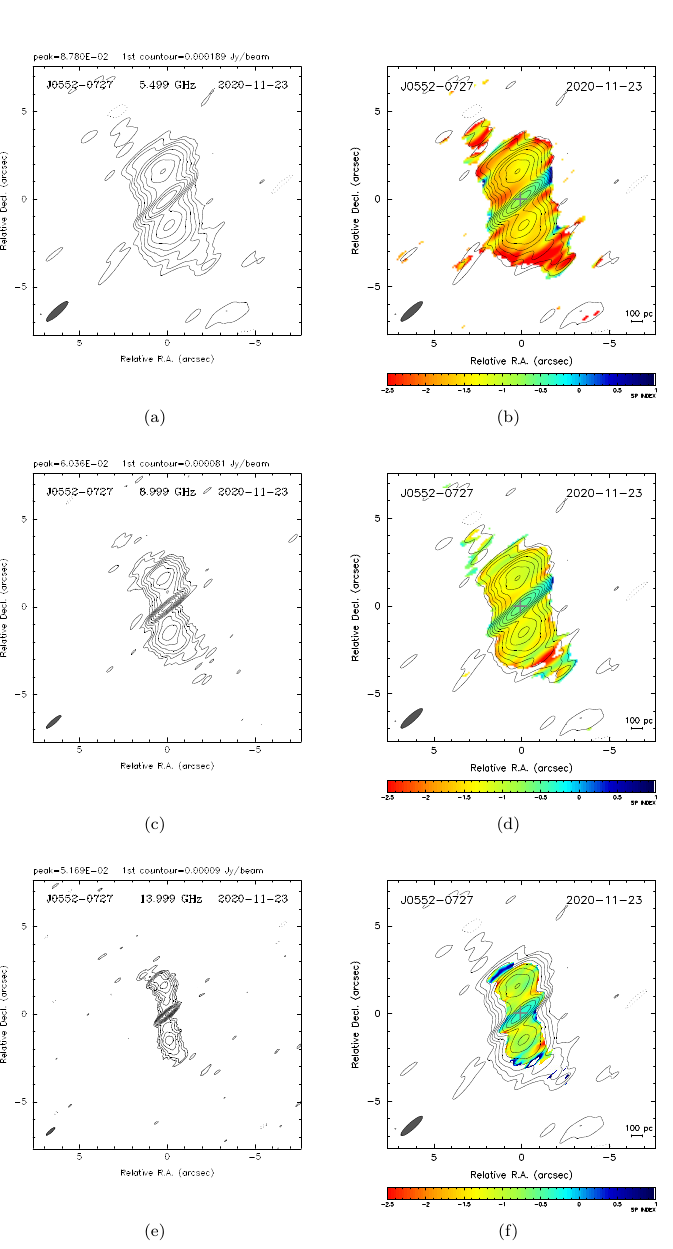} \hfill
    \caption{VLA images of J0552-0727, the second lowest redshift CSO ever confirmed. Subfigures a, c, and e are 5.5, 9, and 14 GHz Stokes I images. Subfigures b, d, and f are 5.5-9, 5.5-14, and 9-14 GHz spectral index maps, respectively. All subfigures have contours overlaid starting at three times the RMS noise level. Grey crosses identify the core.}
    \label{j0552maps_a}
\end{figure}

\begin{figure}[h]
    \centering
    \includegraphics[page=38,trim={2in 4.7in 2in 2in},clip,width=0.68\linewidth]{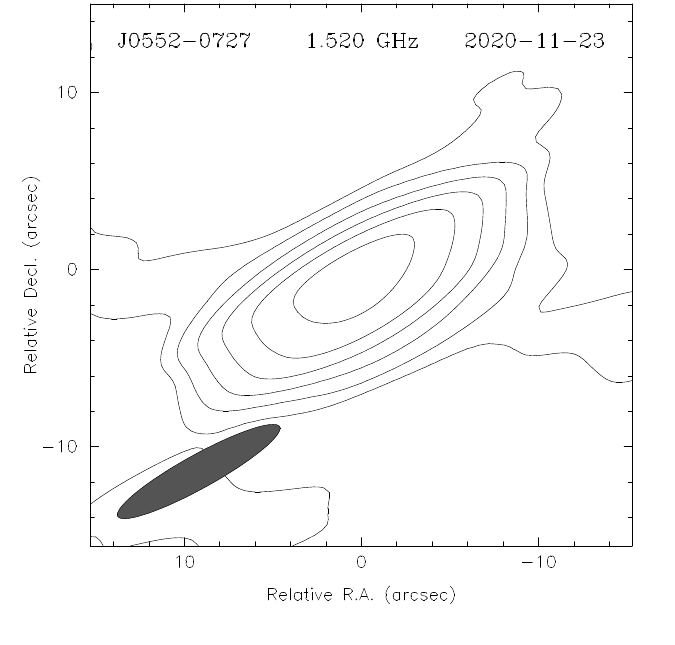} \hfill
    \caption{1.5 GHz Stokes I VLA image of J0552-0727. The lowest contours are at three times the RMS noise level. The boundaries of the map are twice that of the maps in Figure \ref{j0552maps_a}, centered on the same coordinates.}
    \label{j0552maps_b}
\end{figure}

We want to especially highlight this source because it is the second lowest redshift CSO ever confirmed after J1220+2916 (\citetalias{Paper1}). Unlike J1220+2916 however, all of its arcsecond-scale extended emission is resolved out by the VLBA in the 5-15 GHz range. It was only through the VLA observations that we confirmed this source was a CSO. This also means it is one of the few CSOs whose current epoch of emission is observable without the VLBA. All of our Stokes I and spectral index maps of this source are included in Figures \ref{j0552maps_a} and \ref{j0552maps_b}, excluding spectral index maps containing 1.5 GHz because at that frequency, our image has high RMS noise and the source is only barely resolved. The pixel sizes for these maps are 300 mas at 1.5 GHz, 90 mas at 5.5 GHz, 50 mas at 9 GHz, and 30 mas at 14 GHz. \par
We measure the angular size of J0552-0727 to be 5997 $\pm$ 485 mas and its linear size to be 0.978 $\pm$ 0.079 kpc, making it within 1$\sigma$ from exceeding our CSO size cutoff. We assessed its size using the VLA 9 GHz map instead of the 5.5 GHz map for two reasons: because our 9 GHz map has lower RMS noise and because the source has such a low redshift (0.008039), which makes the observed frequency nearly the same as the rest frequency, allowing us to assume that we are still probing the full observable extent of the jet emission. Its size puts it on the borderline between CSOs and medium symmetric objects (MSOs), or symmetrical sources that are between 1 and 20 kpc \citep{Fanti_1995}. Because it is borderline, and there are a range of redshifts reported on NED, we estimated using the highest redshift listed (from \citealt{Jones_2004}, \citealt{Jones_2009}) in order to err on the side of overestimating its size. We classify it as a CSO-1 because of its edge-dimmed lobes. \par

We find that all three components (the core and two lobes) peak below 4.935 GHz. The R$^2$ values for their fits are 0.935, 0.891, and 0.954 for the northern lobe, core, and southern lobe, respectively, exceeding our R$^2$ cutoff for believability in the lobes but not the core. Unfortunately, that means we can only place an upper limit of 4.935 GHz on its turnover frequency.

\subsection{J0650+6001}
We detect an inverted spectrum core in all our spectral index maps of this source, but it is only resolved from the phase center component at 8 and 15 GHz, resulting in four resolved components. Its lobes are edge-dimmed, making it a CSO-1. It also has a high peak luminosity for a CSO-1. This source is included in the \citetalias{CJ1} flux complete sample. However, since it is a CSO-1, we cannot include it in the same statistical analyses conducted on CSO-2s described in \citetalias{Paper2} (see Section \ref{sizedists}).
\subsection{J0906+4636}
This is an archetypical example of a CSO-1. It has a bright, very inverted spectrum core at the phase center with steep spectrum edge-dimmed lobes propagating to the north and south. Like most CSO-1s, it has a redshift below 0.1.
\subsection{J0907+6815} \label{j0907_indiv}
This source would likely not be identifiable as a CSO if we did not have 15 GHz observations of it. Looking at the 5-8 GHz spectral index map, it appears that the southeastern lobe is flat spectrum with a bit of steep spectrum emission a bit further along the jet axis that barely reaches three contour levels. However, at 15 GHz we detect the extended emission out to four contour levels and see that the southeastern hotspot is very compact, indicating that it could be a core component. We are left with two possible interpretations: one is that the southeastern hotspot is the core, making this a CSO-2.1. The other is that the core is unresolved between the two lobes and this is a CSO-2.0. We gave this the final classification of CSO-2.1, though we do not claim to know where the core is.
\subsection{J0909+1928}
This source has two edge-dimmed lobes around a flat-spectrum core, making it a CSO-1. The southern lobe veers off to the southeast while the northern lobe is much dimmer, detected at three contour levels in the 5 GHz map, possibly indicating Doppler boosting. The curve in the southern lobe could be due to an interaction with surrounding gases or jet precession. 
\subsection{J0935+0719}
\begin{figure}[h]
    \centering
    \includegraphics[page=62,trim={5.1cm 3.4cm 5.1cm 4cm},clip,width=0.68\linewidth]{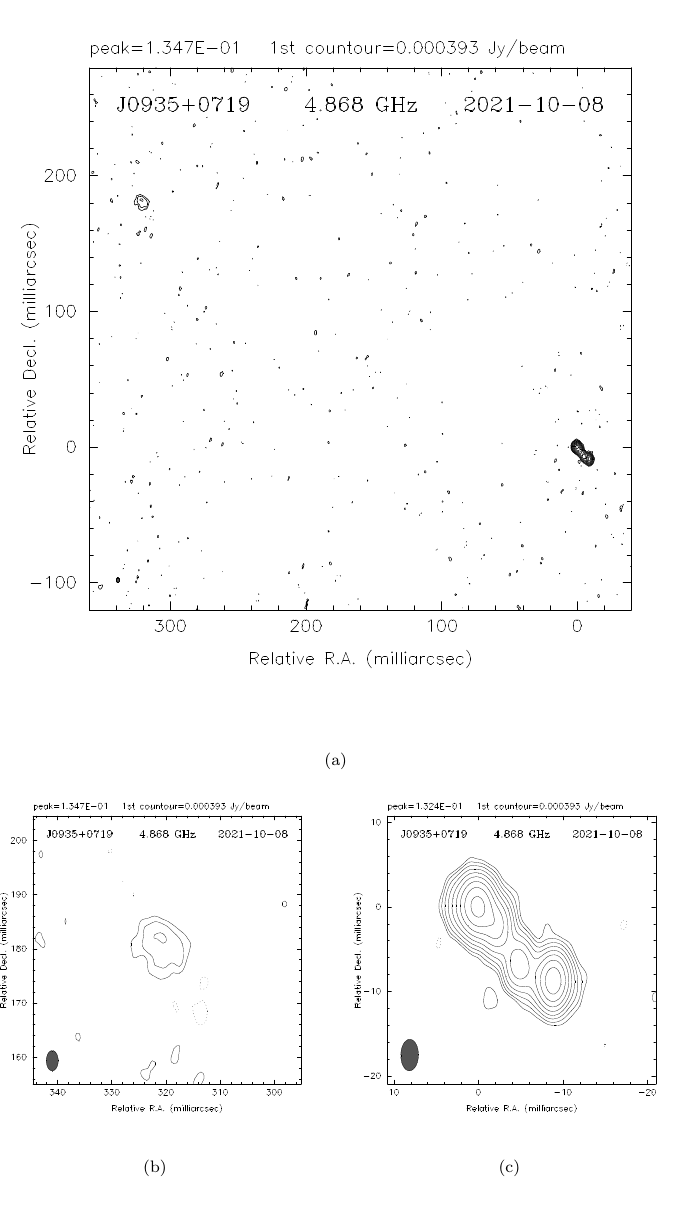}
    \caption{5 GHz VLBA images of J0935+0719, showing the full map on top, extended emission in the bottom left, and the current epoch of emission in the bottom right. The lowest contours are at three times the RMS noise level.}
    \label{j0935_large}
\end{figure}

This source has a very inverted spectrum core that is surrounded on either side by edge-brightened steep spectrum emission, making it a CSO-2.0. There is also some steep spectrum emission at about (-300 mas, +200 mas) in relative (RA, Dec) only detected in the 5 GHz image (Figure \ref{j0935_large}) that does not have a counterpart on the other side of the core. We suspect this is from an earlier epoch of emission since there is no continuous jet leading to the component and it does not have a sharply dimmed outer edge that would imply it is currently being powered. This allows us to keep this as a CSO and not reject it for being over 1 kpc.

\subsection{J1052+8317}
We identify the core of this source as the component at the phase center. Even though it is not very flat spectrum, it is the flattest component in the map and it remains bright and compact at 8 GHz. Because its lobes are compact and edge-brightened, it is a CSO-2.0. Also, since its lobe emission fades away quickly, it has one of the highest core fractions of all 2.0s.
\subsection{J1203+4632}
This source has two steep spectrum lobes, one of which is quite elongated. The bright lobe is much shorter than its diffuse lobe, which is the opposite of what is expected in arm-length ratio studies of AGN. The core's location is atypical; based off of traditional S-morphology common among CSOs, we would expect for it to be the component directly southwest of the flat spectrum component, at approximately (-5 mas, +5 mas) in relative (RA, Dec). In addition, the two lobes appear to trail behind the core to the west at close to a 90$^{\circ}$ from each other, making it a WAT-like CSO and suggesting that the core is moving eastward. The southernmost component bends perpendicularly slightly to the east, which could indicate the jet ramming into dense gases or that the jets are from two different centers of emission. Because of the latter possibility, this source could be an SBBH candidate. We classify this source as a CSO-2.1 because the lobe to the north is more diffuse like a CSO-2.2 and the lobe to the south is brighter and edge-dimmed like a CSO-1. Because of the stark contrast in the flux densities of its lobes, this source may be heavily Doppler boosted and should be investigated further for signs of variability.
\subsection{J1205+2031}
This source has a compact inverted-spectrum core at the phase center and lobes with very different spatial extents. In fact, its morphology appears similar to a core-jet, but it makes the cut as a CSO because of some third contour level emission from its western lobe, which is corroborated by EVN observations by \citet{Cheng_2021} and Radio Fundamental Catalog (RFC, \citealt{Petrov_2024}) images. We classify it as a CSO-2.1 because its eastern lobe is diffuse and its western one is edge-dimmed. Its quickly fading lobes make this the CSO-2.1 with the highest core fraction. As a disclaimer, in order to properly modelfit its core in Difmap with a circular Gaussian, we had to lock its position after establishing it with a few iterations. Otherwise, we kept converging on a solution that split the inverted component's fit between two circular Gaussians approximately equidistant from the phase center. This may indicate that we don't have the whole picture when it comes to this source's morphology, requiring higher angular resolution. 
\subsection{J1215+1730}
This source has a strong inverted spectrum compact core located about (-60 mas,-50 mas) in relative (RA, Dec) from the phase center. Around it are two diffuse steep spectrum lobes, neither of which have very prominent hotspots. Therefore, this is an archetypical CSO-2.2. It has a much lower integrated flux density in our VLBA observations than in our VLA ones, implying that there may be much more lobe emission that the VLBA is resolving out.

\subsection{J1256+5652}
We pinpoint this source's core at the phase center. Surrounding the core are two lobes that are abnormally dim, barely being detected even at 5 GHz. However, we verify their existence with VLBI images at lower frequencies (\citealt{Harding_1999}, RFC). They are edge-brightened, making this source is a 2.0. Unsurprisingly, this source has one of the highest core fractions in our sample. The lobes are most likely dim because they are being resolved out due to their large angular extent, which would be expected given the source's relatively low redshift. With greater (u,v) coverage at smaller spatial scales, we could resolve its low brightness lobe emission and obtain a more accurate core fraction.
\subsection{J1258+5421}
Two flat spectrum components are detected: one at the phase center and another that appears starting at 8 GHz between the phase center and the northeast lobe. Since the second component is only a two contour level detection at only 8 and 15 GHz, we cannot verify its validity and therefore identify the phase center component as the core. This source has a large lobe flux density ratio, indicating possible Doppler boosting. We classify this source as a CSO-1 due to both lobes being edge-dimmed. It also has the highest redshift of any CSO-1 by far at z = 2.65277 (with the next highest at z = 1.12), is an outlier in terms of its redshift versus core fraction (Figure \ref{zcorefrac}), and has a very high peak luminosity for its size, comparable to CSO-2.0s. 
\subsection{J1312+5548}
This source's western jet curves sharply at the end, possibly indicating interactions with the IGM or jet precession. We suspect the core is at the southeastern end of the central ``L" component at $\sim$10 relative RA, giving this source a rather weak counter jet. Since the southeastern lobe is compact with a strong hotspot and the other is so strongly curved away from the jet axis, we classify this as a CSO-2.1.
\subsection{J1340-0335}
We identify the core as the compact component at the phase center; although it is not flat/inverted spectrum, it is the flattest part of the map and stays bright and compact at 15 GHz. The lobes clearly propagate at an angle from each other, making this source a WAT-like CSO due to that angle being larger than 90$^{\circ}$. The eastern lobe component falls off faster than the other, which could indicate Doppler boosting. We label it a CSO-1 due to its edge-dimmed lobes.
\subsection{J1507+5857}
This source has drastically more flux density in its eastern lobe than its western one. In fact, this source would likely be labeled a core-jet were it not for a three contour level detection of the counter jet at 5 GHz. The eastern lobe terminates in a flat spectrum hotspot with heavy perpendicular emission. Paired with the compact western lobe, that makes this source a CSO-2.1. The dramatic difference in lobe flux density could be indicative of Doppler boosting or heavy dispersion due to IGM interactions in the eastern lobe.
\subsection{J1632+3547}
This source has a flat spectrum compact core component at the phase center, most visible in the 5-8 GHz map, that stays bright at high frequencies. It is an example of a WAT-like CSO since its steep spectrum jets emerge at close to 90$^{\circ}$ from each other. The lobes being edge-dimmed also make it a CSO-1.
\subsection{J1639+8631}
This source has a large angular size and is very dim and diffuse, which makes it hard to classify. We are unsure of the core location, but it is likely either at the phase center or undetected just northeast of it. A core detection at the phase center would make this source a WAT-like CSO, but since we don't have definitive evidence, we don't consider it a WAT-like CSO candidate. In the undetected core scenario, this source would be exhibiting classical S-symmetry. There are two long steep-spectrum lobe structures that are misaligned with each other, so we consider this source an SBBH candidate. Because the northern lobe is much more diffuse than the southern one, we classify this source as a CSO-2.1.
\subsection{J1755+6236}
This source exhibits a very inverted spectrum core located at the phase center straddled by two diffuse lobes with complex spectral indices. Its western lobe is somewhat compact however; this makes it a CSO-2.1. Its peak fluxes at all three frequencies are quite low relative to other sources observed in this work. This paired with its diffuse lobe emission makes it our newly confirmed CSO with the lowest VLBA/VLA flux density ratio. Therefore, we are likely resolving out a good portion of its extended emission by observing it with the VLBA. 
\subsection{J1815+6127}
This source features a flat spectrum core situated at the phase center with two jets propagating non-coaxially. Because the jets are between $90^{\circ}$ and $180^{\circ}$ from each other, it is a WAT-like CSO. Both lobes are edge-brightened but diffuse, making this a CSO-2.2.
\subsection{J1845+3541}
We identify the core as the inverted spectrum northeastern edge of the southwestern component that barely resolves some extended emission at 15 GHz. On either side of it are edge-brightened steep spectrum jets that terminate in two fairly compact lobes, making this a CSO-2.0. This source was observed in CJ1 but not verified as a CSO until this paper. We were therefore able to redo some statistical analyses conducted in \citetalias{Paper2}, including it in the list of CSO-2s observed in flux complete samples (see Section \ref{sizedists}).
\subsection{J1855+3742}
This source has an inverted core located slightly south of the phase center around which it hosts two edge-dimmed lobes, making it a CSO-1. It is unique in that it is a CSO-1 with a fairly high redshift of 1.12. It also has the lowest core fraction of all newly confirmed CSO-1s, at 0.0641.
\subsection{J2242+8224}
This source's core is the inverted spectrum component located at the western tip of the bright component on the far west of the large L-shaped region. At 8 GHz, we start to resolve the flat spectrum component in the 5 GHz map into two separate components, the right of which we think is the core. Both this source's lobes are edge-dimmed, making this a CSO-1. This source has high flux ratio lobes where the counter jet is only significantly detected at 5 GHz. We also barely resolve some emission to the north above the counter jet at 5 GHz, which has been corroborated with S band (2-4 GHz) images from the RFC.
\subsection{J2330+3155}
This source appears to have two axes of emission with a possible core component to match each one. One axis extends almost horizontally at $\sim$0 relative Dec across the map with a flat spectrum core at the phase center and edge-dimmed lobes on each side. The other axis lies northeast to southwest with a more inverted core just northeast of the first one. Both of its lobes are compact, with its southwestern one being much dimmer. Due to the two possible axes of emission we consider this source an SBBH candidate. Because of its unclear morphology, we refrain from issuing this source a CSO classification (i.e., CSO-1, CSO-2.0, CSO-2.1, or CSO-2.2).
\newpage

\section{Bona Fide CSOs} \label{bonafidesection}
Table \ref{bonafides} contains a list of the 65 bona fide CSOs newly confirmed in this paper, plus the three bona fide CSOs originally confirmed in \citetalias{Paper1} that were newly observed in this work, totaling 68 sources. Included are the parameters described in Section \ref{csoparams}. Table \ref{bonafidefluxes} contains VLBA peak fluxes and integrated flux densities of the 68 sources at all three frequencies.

Table \ref{bonafides} Redshift References (Col 4): $^1$\citet{Marcha_1996}, $^2$\citet{Fanti_2001}, $^3$\citet{Francis_2000}, $^4$\citet{deVries_2007}, $^5$\citet{Wright_1983}, $^6$\citet{Healey_2008}, $^7$\citet{Jones_2009}, $^8$\citet{Stickel_1993}, $^9$\citet{Glikman_2007}, $^{10}$\citet{Fanti_2011}, $^{11}$\citet{Greene_2007}, $^{12}$\citet{Ahumada_2020}, $^{13}$\citet{Buchanan_2006}, $^{14}$\citet{Albareti_2017}, $^{15}$\citet{Huchra_2012}, $^{16}$\citet{Carilli_1998}, $^{17}$\citet{Henstock_1997}, $^{18}$\citet{Truebenbach_2017}, $^{19}$\citet{Moran_1996}, $^{20}$\citet{Vermeulen_1995}, $^{21}$\citet{Vermeulen_1996}, $^{22}$\citet{Sowards-Emmerd_2005}, $^{23}$\citet{Veron-Cetty_2010}, $^{24}$\citet{Snellen_1996}, $^{25}$\citet{Yee_1996}, $^{26}$\citet{Strader_2014}\\

Table \ref{bonafides} CSO References (Col 11):
$^1$\citet{Dallacasa_2002}, $^2$\citet{Yan_2016}, $^3$\citet{Beasley_2002}, $^4$\citet{Sokolovsky_2011}, $^5$\citet{Taylor_2003}, $^6$\citet{Orienti_2006}, $^7$\citet{Gugliucci_2005}, $^8$\citet{Peck_2000}, $^9$\citet{Augusto_2006}, $^{10}$\citet{Helmboldt_2007}, $^{11}$\citet{Orienti_2007}, $^{12}$\citet{Tremblay_2016}, $^{13}$\citet{Marecki_2014}, $^{14}$\citet{Tremblay_2011}, $^{15}$\citetalias{Paper1}, $^{16}$\citet{Augusto_1996}, $^{17}$\citet{Lonsdale_2003}, $^{18}$\citet{Marr_2014}, $^{19}$\citet{An_2012}, $^{20}$\citet{Xiang_2006}, $^{21}$\citet{Xiang_2005}, $^{22}$\citet{Stanghellini_1997}, $^{23}$This paper

\begin{longrotatetable}
\begin{deluxetable}{ccccccccccc}
\tabletypesize{\footnotesize}
\tablecaption{Data from bona fide CSOs newly observed in this paper. For flux values at each frequency band and core fractions, see Table \ref{bonafidefluxes}. (1) J2000 Name; (2,3) Right Ascension and Declination (h:m:s, $^{\circ}$:$^{\prime}$:$^{\prime \prime}$); (4) Spectroscopic redshift cited from the literature with reference; (5) Largest angular extent (mas), measured this paper; (6) Largest linear size (kpc); (7,8) Frequency and Flux Density at the source's spectral turnover (GHz, Jy). Upper and lower limits are reported for inconclusive plots with an R$^2 \geq$ 0.9. No value is reported if R$^2 <$ 0.9; (9) R$^2$ of power law fit of 0.3385-15 GHz VLA spectrum. No value is reported if R$^2 <$ 0.9, (10) CSO Class. NC indicates CSOs left unclassified due to inconclusive maps; (11) Other references that mention this source being a CSO. A $^\dag$ indicates this source is labeled as a \citetalias{Paper1} CSO in all plots, but uses data newly gathered in this paper.}
\tablehead{\colhead{J2000 Name} & \colhead{RA [h:m:s]} & \colhead{Dec [$^{\circ}$:$^{\prime}$:$^{\prime \prime}$]} & \colhead{Redshift} & \parbox[c]{1.5cm}{\centering Ang. Size [mas]} & \parbox[c]{1.5cm}{\centering Lin. Size [kpc]} & \parbox[c]{1.5cm}{\centering Turnover Freq. [GHz]} & \parbox[c]{1.5cm}{\centering Turnover Flux Dens. [Jy]} & \colhead{R$^2$} & \colhead{CSO Class} & \parbox[c]{1.5cm}{\centering CSO Reference}\\(1)&(2)&(3)&(4)&(5)&(6)&(7)&(8)&(9)&(10)&(11)}
\startdata
J0015$-$1807 & 00h15m34.3244s & $-$18$^{\circ}$07$^{\prime}$25.582$^{\prime \prime}$ & ... & 46.2(2.7) & ... & 1.052(72) & 0.374(7) & 0.971 & 2.0 & 23\\
J0038+2303 & 00h38m08.1014s & +23$^{\circ}$03$^{\prime}$28.5628$^{\prime \prime}$ & 0.096$^1$ & 26.7(3.3) & 0.047(6) & ... & ... & ... & 2.0 & 23\\
J0042+3739 & 00h42m07.1922s & +37$^{\circ}$39$^{\prime}$37.6217$^{\prime \prime}$ & 1.006$^2$ & 109.4(3.8) & 0.882(30) & $\leq$0.339 & $\geq$1.433 & 0.989 & 2.1 & 1\\
J0105+5125 & 01h05m29.5585s & +51$^{\circ}$25$^{\prime}$46.5808$^{\prime \prime}$ & ... & 21.6(3.9) & ... & 1.133(67) & 0.548(9) & 0.979 & 1 & 23\\
J0108$-$1200 & 01h08m13.1724s & $-$12$^{\circ}$00$^{\prime}$50.6373$^{\prime \prime}$ & 1.539$^4$ & 23.7(4.3) & 0.203(37) & ... & ... & ... & 2.0 & 23\\
J0134+0003 & 01h34m12.7042s & +00$^{\circ}$03$^{\prime}$45.1383$^{\prime \prime}$ & 0.879$^3$ & 26.1(2.4) & 0.203(19) & ... & ... & ... & 2.1 & 2,3 \\
J0205+7522 & 02h05m37.9099s & +75$^{\circ}$22$^{\prime}$08.2387$^{\prime \prime}$ & ... & 52.5(3.6) & ... & 0.559(53) & 1.375(55) & 0.977 & 2.0 & 23\\
J0207+6246 & 02h07m03.0168s & +62$^{\circ}$46$^{\prime}$12.0677$^{\prime \prime}$ & ... & 28.3(3.4) & ... & 2.227(82) & 1.867(20) & 0.968 & 2.0 & 4\\
J0210$-$2213 & 02h10m10.058s & $-$22$^{\circ}$13$^{\prime}$37.0025$^{\prime \prime}$ & 1.491$^4$ & 37.6(3.7) & 0.321(31) & 1.459(65) & 0.886(12) & 0.976 & 2.0 & 23\\
J0242$-$2132 & 02h42m35.9099s & $-$21$^{\circ}$32$^{\prime}$25.9347$^{\prime \prime}$ & 0.314$^5$ & 48.9(5.3) & 0.223(24) & $\leq$1.04 & $\geq$1.218 & 0.964 & 1 & 5\\
J0304+7727 & 03h04m54.4442s & +77$^{\circ}$27$^{\prime}$31.6172$^{\prime \prime}$ & ... & 18.9(3.2) & ... & 1.325(60) & 0.761(12) & 0.925 & 2.0 & 4\\
J0401$-$2921 & 04h01m21.4836s & $-$29$^{\circ}$21$^{\prime}$26.829$^{\prime \prime}$ & 0.656$^4$ & 14.8(2.8) & 0.103(20) & $\leq$0.339 & $\geq$1.397 & 0.959 & 2.0 & 23\\
J0428+3259 & 04h28m05.8088s & +32$^{\circ}$59$^{\prime}$52.0443$^{\prime \prime}$ & 0.476$^6$ & 10.1(3.1) & 0.06(2) & 6.75(21) & 0.439(4) & 0.96 & NC & 6\\
J0429+3319 & 04h29m52.7212s & +33$^{\circ}$19$^{\prime}$01.8586$^{\prime \prime}$ & ... & 15.0(3.9) & ... & 2.396(91) & 0.575(7) & 0.947 & 2.0 & 4 \\
J0552$-$0727 & 05h52m11.3762s & $-$07$^{\circ}$27$^{\prime}$22.5182$^{\prime \prime}$ & 0.008039$^7$ & 5996.5(485.1)\tablenotemark{a} & 0.978(79) & $\leq$4.935 & ... & ... & 1 & 23\\ 
J0620+2102 & 06h20m19.5285s & +21$^{\circ}$02$^{\prime}$29.5468$^{\prime \prime}$ & ... & 44.5(4.9) & ... & 1.132(87) & 0.967(15) & 0.989\tablenotemark{b} & 2.0 & 4,7,8\\
J0650+6001 & 06h50m31.2543s & +60$^{\circ}$01$^{\prime}$44.5547$^{\prime \prime}$ & 0.455$^8$ & 12.5(3.2) & 0.072(18) & 3.713(74) & 0.955(11) & 0.928 & 1 & 6,9\\
J0744$-$0629 & 07h44m21.6564s & $-$06$^{\circ}$29$^{\prime}$35.9142$^{\prime \prime}$ & ... & 157.4(2.8) & ... & ... & ... & ...\tablenotemark{b} & 2.2 & 23\\
J0817+1958 & 08h17m05.4948s & +19$^{\circ}$58$^{\prime}$42.931$^{\prime \prime}$ & 0.138$^9$ & 28.6(4.3) & 0.069(10) & $\leq$0.339 & $\geq$0.358 & 0.983 & 1 & 10\\ 
J0843+4215 & 08h43m31.639s & +42$^{\circ}$15$^{\prime}$29.498$^{\prime \prime}$ & 1.0$^{10}$ & 115.4(3.5) & 0.929(28) & $\leq$0.339 & $\geq$2.68 & 0.996 & 2.1 & 1,9,11\\
J0906+4636 & 09h06m15.5367s & +46$^{\circ}$36$^{\prime}$19.028$^{\prime \prime}$ & 0.085$^{11}$ & 33.1(3.6) & 0.052(6) & ... & ... & ... & 1 & 12\\
J0907+6815 & 09h07m52.9463s & +68$^{\circ}$15$^{\prime}$44.920$^{\prime \prime}$ & ... & 16.0(3.5) & ... & 2.839(92) & 0.245(3) & 0.973 & 2.1 & 23\\
J0909+1928 & 09h09m37.4414s & +19$^{\circ}$28$^{\prime}$08.297$^{\prime \prime}$ & 0.027843$^{12}$ & 17.6(4.5) & 0.01(0) & ... & ... & ... & 1 & 15$^{\dag}$\\
J0935+0719 & 09h35m01.076s & +07$^{\circ}$19$^{\prime}$18.6105$^{\prime \prime}$ & 0.29$^{13}$ & 21.5(4.4) & 0.093(19) & 1.866(87) & 0.521(6) & 0.995 & 2.0 & 13 \\ 
J1006+4836 & 10h06m39.5762s & +48$^{\circ}$36$^{\prime}$31.196$^{\prime \prime}$ & 0.405113$^{14}$ & 10.5(3.3) & 0.057(18) & 4.0(1) & 0.119(1) & 0.966 & 2.0 & 10,12\\
J1025+1022 & 10h25m44.230s & +10$^{\circ}$22$^{\prime}$30.52$^{\prime \prime}$ & 0.045805$^{15}$ & 23.3(3.8) & 0.021(3) & ... & ... & ... & 1 & 15$^{\dag}$\\
J1052+8317 & 10h52m18.2076s & +83$^{\circ}$17$^{\prime}$26.6307$^{\prime \prime}$ & ... & 32.0(3.3) & ... & 1.13(9) & 0.197(4) & 0.969 & 2.0 & 23\\ 
J1143+1834 & 11h43m26.0703s & +18$^{\circ}$34$^{\prime}$38.373$^{\prime \prime}$ & ... & 16.2(4.7) & ... & 3.095(81) & 0.434(5) & 0.926 & 2.0 & 4,7,8,10,12\\
J1203+4632 & 12h03m31.7958s & +46$^{\circ}$32$^{\prime}$55.562$^{\prime \prime}$ & 0.606067$^{14}$ & 40.7(3.7) & 0.273(25) & $\leq$1.04 & $\geq$0.474 & 0.928 & 2.1 & 2,10,14\\ 
J1205+2031 & 12h05m51.465s & +20$^{\circ}$31$^{\prime}$19.07$^{\prime \prime}$ & 0.02378857$^{12}$ & 24.4(3.6) & 0.012(2) & ... & ... & ... & 2.1 & 15$^{\dag}$\\
J1215+1730 & 12h15m14.7218s & +17$^{\circ}$30$^{\prime}$02.238$^{\prime \prime}$ & 0.176275$^{14}$ & 139.8(3.7) & 0.413(11) & $\leq$0.339 & $\geq$2.154 & 0.998 & 2.2 & 2,12,16\\
J1240+2405 & 12h40m47.984s & +24$^{\circ}$05$^{\prime}$14.188$^{\prime \prime}$ & 3.58705$^{14}$ & 51.1(4.6) & 0.378(34) & $\leq$0.339 & $\geq$0.777 & 0.991 & 2.0 & 10,12 \\
J1256+5652 & 12h56m14.2345s & +56$^{\circ}$52$^{\prime}$25.235$^{\prime \prime}$ & 0.04217$^{16}$ & 52.5(3.4) & 0.043(3) & ... & ... & ... & 2.0 & 10,17\\
J1258+5421 & 12h58m15.6094s & +54$^{\circ}$21$^{\prime}$52.125$^{\prime \prime}$ & 2.65277$^{14}$ & 64.4(3.2) & 0.521(26) & $\leq$0.339 & $\geq$0.998 & 0.996 & 1 & 10 \\ 
J1312+5548 & 13h12m53.1973s & +55$^{\circ}$48$^{\prime}$13.098$^{\prime \prime}$ & 1.083992$^{12}$ & 27.7(4.1) & 0.227(34) & 0.404(42) & 0.713(36) & 0.938 & 2.1 & 10 \\ 
J1324+4048 & 13h24m12.0974s & +40$^{\circ}$48$^{\prime}$11.766$^{\prime \prime}$ & 0.495$^{17}$ & 15.3(4.6) & 0.093(28) & 3.02(9) & 0.385(4) & 0.983 & 2.0 & 4,10,12,18,19 \\
J1326+5712 & 13h26m50.5718s & +57$^{\circ}$12$^{\prime}$06.759$^{\prime \prime}$ & 0.459042$^{14}$ & 14.0(3.4) & 0.081(20) & ... & ... & ... & 1 & 23\\
J1340$-$0335 & 13h40m13.304s & $-$03$^{\circ}$35$^{\prime}$20.803$^{\prime \prime}$ & ... & 18.7(2.7) & ... & 1.146(82) & 0.89(2) & 0.99 & 1 & 23\\
J1442+3042 & 14h42m41.5312s & +30$^{\circ}$42$^{\prime}$32.941$^{\prime \prime}$ & 0.542963$^{14}$ & 42.5(4.3) & 0.27(3) & 0.774(68) & 0.492(15) & 0.995 & 2.0 & 10,12\\
J1451+1343 & 14h51m31.4909s & +13$^{\circ}$43$^{\prime}$24.0001$^{\prime \prime}$ & ... & 50.1(3.0) & ... & 0.458(48) & 0.904(43) & 0.98 & 1 & 4\\
J1507+5857 & 15h07m47.3894s & +58$^{\circ}$57$^{\prime}$27.647$^{\prime \prime}$ & 0.31$^{18}$ & 38.3(3.0) & 0.173(14) & $\leq$0.339 & $\geq$0.623 & 0.984 & 2.1 & 10\\ 
J1537+8154 & 15h37m00.0373s & +81$^{\circ}$54$^{\prime}$30.9935$^{\prime \prime}$ & ... & 17.4(3.6) & ... & 0.631(53) & 0.45(1) & 0.945 & 1 & 23\\
J1632+3547 & 16h32m31.2544s & +35$^{\circ}$47$^{\prime}$37.736$^{\prime \prime}$ & ... & 19.0(4.1) & ... & 1.126(70) & 0.522(10) & 0.977 & 1 & 23\\ 
J1639+8631 & 16h39m25.0218s & +86$^{\circ}$31$^{\prime}$53.1389$^{\prime \prime}$ & ... & 72.6(3.3) & ... & ... & ... & ... & 2.1 & 23\\
J1700+3830 & 17h00m19.9678s & +38$^{\circ}$30$^{\prime}$34.14$^{\prime \prime}$ & 0.582312$^{14}$ & 52.8(3.1) & 0.348(20) & 0.721(70) & 0.489(17) & 0.968 & 2.0 & 10,12 \\
J1730+3811 & 17h30m54.1146s & +38$^{\circ}$11$^{\prime}$50.87$^{\prime \prime}$ & ... & 36.4(3.3) & ... & $\leq$0.339 & $\geq$0.739 & 0.992 & 2.0 & 10,12\\
J1755+6236 & 17h55m48.4407s & +62$^{\circ}$36$^{\prime}$44.118$^{\prime \prime}$ & 0.027$^{19}$ & 55.7(3.3) & 0.03(0) & $\leq$0.339 & $\geq$0.419 & 0.989\tablenotemark{b} & 2.1 & 23\\
J1815+6127 & 18h15m36.7922s & +61$^{\circ}$27$^{\prime}$11.6477$^{\prime \prime}$ & 0.601$^{20}$ & 17.2(3.2) & 0.115(22) & 0.62(6) & 0.901(35) & 0.984 & 2.2 & 23\\
J1819$-$0258 & 18h19m17.4085s & $-$02$^{\circ}$58$^{\prime}$07.872$^{\prime \prime}$ & ... & 27.4(2.7) & ... & ... & ... & ... & 2.0 & 23\\
J1823+7938 & 18h23m14.1086s & +79$^{\circ}$38$^{\prime}$49.002$^{\prime \prime}$ & 0.224$^{17}$ & 22.8(3.4) & 0.081(12) & 4.299(88) & 0.632(7) & 0.958 & 2.0 & 23\\
J1845+3541 & 18h45m35.1088s & +35$^{\circ}$41$^{\prime}$16.7263$^{\prime \prime}$ & 0.764$^{21}$ & 17.8(3.3) & 0.132(25) & 2.488(76) & 0.953(11) & 0.974 & 2.0 & 9\\
J1855+3742 & 18h55m27.7068s & +37$^{\circ}$42$^{\prime}$56.9659$^{\prime \prime}$ & 1.12$^{22}$ & 14.0(4.3) & 0.116(36) & ... & ... & ... & 1 & 6\\
J1909+7813 & 19h09m18.9584s & +78$^{\circ}$13$^{\prime}$29.8741$^{\prime \prime}$ & ... & 152.1(2.8) & ... & ... & ... & ... & 2.0 & 23\\
J1921+4333 & 19h21m09.9348s & +43$^{\circ}$33$^{\prime}$41.8362$^{\prime \prime}$ & ... & 10.4(2.9) & ... & 3.128(85) & 0.262(3) & 0.96 & 2.0 & 4\\
J1935+8130 & 19h35m22.7234s & +81$^{\circ}$30$^{\prime}$14.5524$^{\prime \prime}$ & ... & 16.0(3.0) & ... & 3.764(85) & 0.483(6) & 0.976 & 2.0 & 4,5\\
J1950$-$0436 & 19h50m44.055s & $-$04$^{\circ}$36$^{\prime}$11.839$^{\prime \prime}$ & ... & 73.3(5.7) & ... & ... & ... & ... & 2.0 & 23\\
J1950+0807 & 19h50m05.54s & +08$^{\circ}$07$^{\prime}$13.9797$^{\prime \prime}$ & ... & 33.1(6.4) & ... & 1.615(75) & 1.244(15) & 0.913 & 2.0 & 4\\
J2010$-$2425 & 20h10m45.153s & $-$24$^{\circ}$25$^{\prime}$45.55$^{\prime \prime}$ & 0.825$^{26}$ & 53.9(2.7) & 0.41(2) & ... & ... & ... & 2.1 & 23\\
J2035+1857 & 20h35m33.9834s & +18$^{\circ}$57$^{\prime}$05.465$^{\prime \prime}$ & ... & 51.5(4.2) & ... & ... & ... & ... & 2.2 & 23\\
J2052+3635 & 20h52m52.055s & +36$^{\circ}$35$^{\prime}$35.3005$^{\prime \prime}$ & 0.355$^{23}$ & 69.8(2.8) & 0.346(14) & 1.713(90) & 4.472(56) & 0.983 & 2.0 & 23\\
J2120+6642 & 21h20m46.2019s & +66$^{\circ}$42$^{\prime}$20.2336$^{\prime \prime}$ & ... & 15.8(3.4) & ... & 3.165(83) & 0.225(3) & 0.981 & 2.0 & 4 \\
J2123$-$0112 & 21h23m39.1467s & $-$01$^{\circ}$12$^{\prime}$34.7086$^{\prime \prime}$ & 1.158$^{24}$ & 71.0(3.5) & 0.589(29) & ... & ... & ... & 2.0 & 20 \\
J2130+0502 & 21h30m32.8774s & +05$^{\circ}$02$^{\prime}$17.4754$^{\prime \prime}$ & 0.99$^8$ & 50.1(5.7) & 0.402(46) & ... & ... & ... & 2.0 & 22\\
J2131+8430 & 21h31m39.3216s & +84$^{\circ}$30$^{\prime}$11.8184$^{\prime \prime}$ & ... & 22.8(3.0) & ... & 0.772(62) & 0.7(0) & 0.976 & 2.2 & 4\\
J2153+1741 & 21h53m36.8261s & +17$^{\circ}$41$^{\prime}$43.6888$^{\prime \prime}$ & 0.23024$^{25}$ & 36.3(5.5) & 0.132(20) & ... & ... & ... & 1 & 21\\
J2242+8224 & 22h42m43.6381s & +82$^{\circ}$24$^{\prime}$47.0106$^{\prime \prime}$ & ... & 33.8(3.2) & ... & $\leq$0.339 & $\geq$0.634 & 0.988 & 1 & 23\\
J2325$-$0344 & 23h25m10.2581s & $-$03$^{\circ}$44$^{\prime}$46.7145$^{\prime \prime}$ & 1.509$^4$ & 53.2(5.8) & 0.455(50) & 1.011(56) & 1.128(21) & 0.927 & 2.0 & 20,21\\
J2330+3155 & 23h30m46.1599s & +31$^{\circ}$55$^{\prime}$33.507$^{\prime \prime}$ & ... & 40.9(4.7) & ... & ... & ... & ... & NC & 23
\enddata
\label{bonafides}
\tablenotetext{a}{Angular size measured from 8 GHz VLA image}
\tablenotetext{b}{15 GHz excluded}
\end{deluxetable}
\end{longrotatetable}

\startlongtable
\begin{deluxetable}{c|cc|ccc|cc}

\tablecaption{Peak flux and flux density values and 8 GHz core fractions for newly observed bona fide CSOs.}
\tablehead{\colhead{Source Name} & \colhead{$S_{5}$ [Jy]} & \parbox[c]{2cm}{\centering $S_{peak,5}$ [${\rm Jy\,beam}^{-1}$]} & \colhead{$S_{8}$ [Jy]} & \parbox[c]{2cm}{\centering $S_{peak,8}$ [${\rm Jy\,beam}^{-1}$]} & \parbox[c]{2cm}{\centering Core Fraction at 8 GHz} & \colhead{$S_{15}$ [Jy]} & \parbox[c]{2cm}{\centering $S_{peak,15}$ [${\rm Jy\,beam}^{-1}$]}}
\startdata
J0015-1807 & 0.255(13) & 0.158(8) & 0.173(9) & 0.110(6) & 0.543(10) & 0.107(5) & 0.055(3) \\ 
J0038+2303 & 0.412(21) & 0.175(9) & 0.220(11) & 0.059(3) & 0.115(11) & 0.109(5) & 0.022(1) \\ 
J0042+3739 & 0.148(7) & 0.020(1) & 0.074(4) & 0.013(1) & $\leq$8.33e-03 & ... & ... \\ 
J0105+5125 & 0.378(19) & 0.177(9) & 0.156(8) & 0.072(4) & 0.432(8) & 0.083(4) & 0.044(2) \\ 
J0108-1200 & 0.279(14) & 0.066(3) & 0.136(7) & 0.027(1) & 0.045(7) & ... & ... \\ 
J0134+0003 & 0.472(24) & 0.208(10) & 0.264(13) & 0.100(5) & 0.138(13) & 0.196(10) & 0.049(2) \\ 
J0205+7522 & 0.667(33) & 0.241(12) & 0.318(16) & 0.101(5) & $\leq$3.25e-03 & 0.133(7) & 0.034(2) \\ 
J0207+6246 & 1.841(92) & 1.200(60) & 0.995(50) & 0.643(32) & $\leq$1.76e-03 & 0.516(26) & 0.266(13) \\ 
J0210-2213 & 0.696(35) & 0.273(14) & 0.296(15) & 0.107(5) & 0.025(15) & 0.107(5) & 0.026(1) \\ 
J0242-2132 & 0.472(24) & 0.163(8) & 0.362(18) & 0.078(4) & 0.406(20) & 0.183(9) & 0.034(2) \\ 
J0304+7727 & 0.476(24) & 0.158(8) & 0.244(12) & 0.065(3) & 0.009(12) & 0.111(6) & 0.031(2) \\ 
J0401-2921 & 0.265(13) & 0.100(5) & 0.248(12) & 0.057(3) & 0.208(13) & 0.130(6) & 0.029(1) \\ 
J0428+3259 & 0.540(27) & 0.391(20) & 0.447(22) & 0.286(14) & 0.599(26) & 0.320(16) & 0.195(10) \\ 
J0429+3319 & 0.653(33) & 0.403(20) & 0.376(19) & 0.207(10) & $\leq$3.12e-03 & 0.203(10) & 0.088(4) \\ 
J0552$-$0727\tablenotemark{a} & 0.088(4) & 0.073(4) & 0.048(2) & 0.041(2) & ... & 0.045(2) & 0.037(2) \\ 
J0620+2102 & 0.541(27) & 0.264(13) & 0.299(15) & 0.142(7) & $\leq$3.09e-03 & 0.138(7) & 0.052(3) \\ 
J0650+6001 & 1.068(53) & 0.617(31) & 0.749(37) & 0.380(19) & 0.036(37) & 0.443(22) & 0.167(8) \\ 
J0744$-$0629\tablenotemark{b} & 1.639(82) & 0.255(13) & 1.068(53) & 0.131(7) & $\leq$1.29e-03 & 0.315(16) & 0.056(3) \\ 
J0817+1958 & 0.161(8) & 0.064(3) & 0.121(6) & 0.038(2) & 0.300(6) & ... & ... \\  
J0843+4215\tablenotemark{c} & 0.451(23) & 0.125(6) & 0.217(11) & 0.037(2) & $\leq$4.67e-03 & ... & ... \\
J0906+4636 & 0.128(6) & 0.081(4) & 0.112(6) & 0.085(4) & 0.688(7) & 0.078(4) & 0.059(3) \\ 
J0907+6815 & 0.268(13) & 0.180(9) & 0.173(9) & 0.103(5) & $\leq$3.68e-03 & 0.076(4) & 0.035(2) \\ 
J0909+1928$^*$ & 0.125(6) & 0.096(5) & 0.103(5) & 0.074(4) & 0.525(6) & 0.079(4) & 0.063(3) \\ 
J0935+0719 & 0.324(16) & 0.132(7) & 0.162(8) & 0.041(2) & 0.138(8) & ... & ... \\ 
J1006+4836 & 0.113(6) & 0.065(3) & 0.088(4) & 0.045(2) & $\leq$7.69e-03 & 0.052(3) & 0.024(1) \\ 
J1025+1022$^*$ & 0.103(5) & 0.038(2) & 0.089(4) & 0.029(1) & 0.429(5) & ... & ... \\ 
J1052+8317 & 0.106(5) & 0.052(3) & 0.049(2) & 0.027(1) & 0.671(3) & ... & ... \\ 
J1143+1834 & 0.322(16) & 0.168(8) & 0.260(13) & 0.133(7) & $\leq$2.58e-03 & 0.124(6) & 0.064(3) \\ 
J1203+4632 & 0.158(8) & 0.061(3) & 0.116(6) & 0.040(2) & 0.399(6) & ... & ... \\ 
J1205+2031$^*$ & 0.037(2) & 0.022(1) & 0.050(3) & 0.032(2) & 0.692(3) & ... & ... \\ 
J1215+1730\tablenotemark{d} & 0.190(10) & 0.016(1) & 0.112(6) & 0.021(1) & 0.192(6) & 0.040(2) & 0.022(1) \\ 
J1240+2405 & 0.227(11) & 0.086(4) & 0.184(9) & 0.048(2) & $\leq$3.84e-03 & 0.110(6) & 0.021(1) \\ 
J1256+5652 & 0.596(30) & 0.526(26) & 0.639(32) & 0.475(24) & 0.759(40) & 0.425(21) & 0.237(12) \\ 
J1258+5421 & 0.198(10) & 0.040(2) & 0.100(5) & 0.030(2) & 0.259(5) & 0.043(2) & 0.024(1) \\ 
J1312+5548 & 0.266(13) & 0.072(4) & 0.121(6) & 0.254(13) & 0.129(6) & ... & ... \\ 
J1324+4048 & 0.422(21) & 0.204(10) & 0.218(11) & 0.101(5) & $\leq$4.61e-03 & 0.086(4) & 0.035(2) \\ 
J1326+5712 & 0.287(14) & 0.139(7) & 0.224(11) & 0.134(7) & 0.575(13) & 0.155(8) & 0.105(5) \\ 
J1340-0335 & 0.229(11) & 0.095(5) & 0.117(6) & 0.041(2) & 0.374(6) & 0.035(2) & 0.026(1) \\ 
J1442+3042 & 0.179(9) & 0.036(2) & 0.119(6) & 0.020(1) & $\leq$5.25e-03 & ... & ... \\ 
J1451+1343 & 0.361(18) & 0.138(7) & 0.204(10) & 0.080(4) & $\leq$4.59e-03 & 0.095(5) & 0.034(2) \\ 
J1507+5857 & 0.235(12) & 0.037(2) & 0.165(8) & 0.034(2) & 0.222(8) & 0.079(4) & 0.022(1) \\ 
J1537+8154 & 0.175(9) & 0.076(4) & 0.130(7) & 0.047(2) & 0.116(7) & ... & ... \\ 
J1632+3547 & 0.273(14) & 0.120(6) & 0.172(9) & 0.083(4) & 0.480(10) & 0.098(5) & 0.046(2) \\ 
J1639+8631 & 0.158(8) & 0.044(2) & 0.099(5) & 0.036(2) & $\leq$6.66e-03 & ... & ... \\ 
J1700+3830 & 0.170(9) & 0.095(5) & 0.073(4) & 0.037(2) & $\leq$1.01e-02 & ... & ... \\ 
J1730+3811 & 0.318(16) & 0.089(4) & 0.125(6) & 0.028(1) & $\leq$8.24e-03 & ... & ... \\ 
J1755+6236 & 0.039(2) & 0.011(1) & 0.035(2) & 0.017(1) & 0.489(2) & ... & ... \\ 
J1815+6127 & 0.676(34) & 0.263(13) & 0.398(20) & 0.192(10) & 0.476(22) & 0.238(12) & 0.135(7) \\ 
J1819-0258 & 0.941(47) & 0.179(9) & 0.568(28) & 0.059(3) & $\leq$1.30e-03 & 0.294(15) & 0.025(1) \\ 
J1823+7938 & 0.780(39) & 0.352(18) & 0.603(30) & 0.241(12) & $\leq$2.12e-03 & 0.308(15) & 0.117(6) \\ 
J1845+3541 & 0.968(48) & 0.399(20) & 0.553(28) & 0.174(9) & 0.011(28) & 0.248(12) & 0.085(4) \\ 
J1855+3742 & 0.425(21) & 0.321(16) & 0.222(11) & 0.123(6) & 0.064(11) & 0.115(6) & 0.040(2) \\ 
J1909+7813\tablenotemark{b} & 0.173(9) & 0.026(1) & 0.067(3) & 0.016(1) & $\leq$9.73e-03 & 0.035(2) & 0.026(1) \\ 
J1921+4333 & 0.230(12) & 0.094(5) & 0.202(10) & 0.068(3) & $\leq$3.59e-03 & 0.120(6) & 0.024(1) \\ 
J1935+8130 & 0.587(29) & 0.309(15) & 0.393(20) & 0.180(9) & $\leq$2.08e-03 & 0.174(9) & 0.068(3) \\ 
J1950-0436 & 0.753(38) & 0.198(10) & 0.295(15) & 0.078(4) & $\leq$2.34e-03 & 0.120(6) & 0.028(1) \\ 
J1950+0807 & 1.409(70) & 0.691(35) & 0.881(44) & 0.403(20) & $\leq$2.04e-03 & 0.457(23) & 0.183(9) \\ 
J2010-2425 & 0.248(12) & 0.034(2) & 0.144(7) & 0.019(1) & 0.206(7) & ... & ... \\ 
J2035+1857 & 0.290(14) & 0.061(3) & 0.145(7) & 0.018(1) & 0.126(7) & 0.027(1) & 0.021(1) \\ 
J2052+3635 & 4.508(225) & 1.306(65) & 1.932(97) & 0.441(22) & $\leq$1.28e-03 & 0.805(40) & 0.132(7) \\ 
J2120+6642 & 0.276(14) & 0.139(7) & 0.158(8) & 0.071(4) & $\leq$5.50e-03 & 0.085(4) & 0.032(2) \\ 
J2123-0112 & 0.371(19) & 0.095(5) & 0.134(7) & 0.020(1) & $\leq$4.74e-03 & ... & ... \\ 
J2130+0502 & 2.067(103) & 0.679(34) & 1.267(63) & 0.278(14) & $\leq$7.10e-04 & 0.662(33) & 0.081(4) \\ 
J2131+8430 & 0.548(27) & 0.224(11) & 0.239(12) & 0.089(4) & $\leq$3.44e-03 & 0.115(6) & 0.027(1) \\ 
J2153+1741 & 0.187(9) & 0.014(1) & 0.112(6) & 0.086(4) & 0.763(7) & 0.058(3) & 0.032(2) \\ 
J2242+8224 & 0.227(11) & 0.042(2) & 0.116(6) & 0.017(1) & 0.064(6) & ... & ... \\ 
J2325-0344 & 0.704(35) & 0.364(18) & 0.301(15) & 0.150(8) & $\leq$2.73e-03 & 0.140(7) & 0.041(2) \\ 
J2330+3155 & 0.732(37) & 0.426(21) & 0.405(20) & 0.268(13) & $\leq$1.98e-03 & 0.229(11) & 0.138(7) 
\label{bonafidefluxes}
\enddata
\tablenotetext{a}{Numbers reported refer to just the core component since that was the only component resolved in the VLBA images. No core fraction reported.}
\tablenotetext{b}{15 GHz Stokes I pixels are 0.15 mas}
\tablenotetext{c}{15 GHz Stokes I pixels are 0.117 mas}
\tablenotetext{d}{15 GHz Stokes I pixels are 0.12 mas}
\tablecomments{Bona fide CSOs originally verified in \citetalias{Paper1} are marked with an asterisk ($^*$). Upper limits given for core fractions represent a 6$\sigma$ threshold, or 2$^\mathrm{nd}$ contour level for detection. No 15 GHz flux values indicates the peak flux was below our 20 m${\rm Jy\,beam}^{-1}$ cutoff. Pixel sizes for the Stokes I maps are 0.25 mas at 5 GHz, 0.2 mas at 8 GHz, and 0.1 mas at 15 GHz unless otherwise stated.}
\end{deluxetable}
\newpage

\section{Rejected A-Candidates} \label{rejectedsection}
\setcounter{table}{0}
This section contains a list of the 56 A-candidates that were refuted as CSOs, including position information and the criterion used to reject each of them.
\startlongtable
\begin{deluxetable}{c|ccc}
    \tablecaption{Refuted A-Candidate Sources}
    \tablehead{\colhead{Source Name} & \colhead{RA} & \colhead{Dec} & \colhead{Rejection Criterion}}
    \startdata
        J0017+5312 & 00h17m51.7598s & +53$^{\circ}$12$^{\prime}$19.1219$^{\prime \prime}$ & M\\
        J0037$-$2145 & 00h37m14.8259s & $-$21$^{\circ}$45$^{\prime}$24.714$^{\prime \prime}$ & M\\
        J0048+3157$^{\dag}$ & 00h48m47.1415s & +31$^{\circ}$57$^{\prime}$25.0849$^{\prime \prime}$ & V\\
        J0048+0640 & 00h48m58.7231s & +06$^{\circ}$40$^{\prime}$06.475$^{\prime \prime}$ & M\\
        J0101$-$2831 & 01h01m52.3897s & $-$28$^{\circ}$31$^{\prime}$20.4284$^{\prime \prime}$ & M\\
        J0119+0829 & 01h19m01.2743s & +08$^{\circ}$29$^{\prime}$54.7046$^{\prime \prime}$ & M\\
        J0128+6306 & 01h28m30.565s & +63$^{\circ}$06$^{\prime}$29.8821$^{\prime \prime}$ & M\\
        J0214$-$2438 & 02h14m55.650s & $-$24$^{\circ}$38$^{\prime}$16.30$^{\prime \prime}$ & M\\
        J0329+2756 & 03h29m57.6694s & +27$^{\circ}$56$^{\prime}$15.499$^{\prime \prime}$ & M\\
        J0347+2004 & 03h47m29.5591s & +20$^{\circ}$04$^{\prime}$53.043$^{\prime \prime}$ & M\\
        J0519+7133 & 05h19m28.8819s & +71$^{\circ}$33$^{\prime}$03.7257$^{\prime \prime}$& M\\
        J0631+5311 & 06h31m34.6853s & +53$^{\circ}$11$^{\prime}$27.7531$^{\prime \prime}$ & M\\
        J0753+4231 & 07h53m03.3385s & +42$^{\circ}$31$^{\prime}$30.761$^{\prime \prime}$& M\\
        J0814$-$1806 & 08h14m07.9008s & $-$18$^{\circ}$06$^{\prime}$26.0543$^{\prime \prime}$& M\\
        J0818+6109 & 08h18m13.61s & +61$^{\circ}$09$^{\prime}$28.501$^{\prime \prime}$& M\\
        J0821$-$0323 & 08h21m40.0376s & $-$03$^{\circ}$23$^{\prime}$12.5387$^{\prime \prime}$ & M\\
        J0909+0835 & 09h09m12.1575s & +08$^{\circ}$35$^{\prime}$41.099$^{\prime \prime}$ & M \\
        J0913+1454 & 09h13m34.9813s & +14$^{\circ}$54$^{\prime}$20.0987$^{\prime \prime}$ & M\\
        J1005+2403 & 10h05m07.8678s & +24$^{\circ}$03$^{\prime}$37.996$^{\prime \prime}$ & M\\
        J1008$-$0933 & 10h08m43.8654s & $-$09$^{\circ}$33$^{\prime}$23.3622$^{\prime \prime}$ & M\\
        J1011+7124\tablenotemark{a,b} & 10h11m32.618075s & +71$^{\circ}$24$^{\prime}$41.59272$^{\prime \prime}$& S\\
        J1024$-$0052 & 10h24m29.5865s & $-$00$^{\circ}$52$^{\prime}$55.498$^{\prime \prime}$ & M\\
        J1032+5610 & 10h32m02.5113s & +56$^{\circ}$10$^{\prime}$56.721$^{\prime \prime}$ & M\\
        J1036$-$0605 & 10h36m47.573s & $-$06$^{\circ}$05$^{\prime}$41.1847$^{\prime \prime}$ & M\\
        J1042+0748 & 10h42m57.5887s & +07$^{\circ}$48$^{\prime}$50.548$^{\prime \prime}$ & M\\
        J1052+3811\tablenotemark{c,d} & 10h52m11.7904s & +38$^{\circ}$11$^{\prime}$44.0173$^{\prime \prime}$& S\\
        J1110+4817 & 11h10m36.3247s & +48$^{\circ}$17$^{\prime}$52.444$^{\prime \prime}$ & M\\
        J1139+3803 & 11h39m34.0095s & +38$^{\circ}$03$^{\prime}$41.9667$^{\prime \prime}$ & M\\
        J1140+5912\tablenotemark{e} & 11h40m49.577s & +59$^{\circ}$12$^{\prime}$25.161$^{\prime \prime}$ & M\\
        J1141+4945 & 11h41m54.8254s & +49$^{\circ}$45$^{\prime}$06.564$^{\prime \prime}$ & M\\
        J1201+3919 & 12h01m49.9663s & +39$^{\circ}$19$^{\prime}$11.038$^{\prime \prime}$ & M\\
        J1210+6422 & 12h10m31.6419s & +64$^{\circ}$22$^{\prime}$17.476$^{\prime \prime}$ & M\\
        J1211$-$1926 & 12h11m57.7383s & $-$19$^{\circ}$26$^{\prime}$07.6597$^{\prime \prime}$ & M\\
        J1225+3914 & 12h25m50.5693s & +39$^{\circ}$14$^{\prime}$22.674$^{\prime \prime}$ & M\\
        J1241+6020 & 12h41m29.5913s & +60$^{\circ}$20$^{\prime}$41.331$^{\prime \prime}$ & M\\
        J1310+3404 & 13h10m04.4335s & +34$^{\circ}$03$^{\prime}$09.088$^{\prime \prime}$ & M\\
        J1313+6735 & 13h13m27.989s & +67$^{\circ}$35$^{\prime}$50.36$^{\prime \prime}$ & M\\
        J1319+3840 & 13h19m59.7758s & +38$^{\circ}$40$^{\prime}$22.43$^{\prime \prime}$ & M\\
        J1320+8450 & 13h20m53.1855s & +84$^{\circ}$50$^{\prime}$11.1547$^{\prime \prime}$ & M\\
        J1344$-$1739 & 13h44m03.420s & $-$17$^{\circ}$39$^{\prime}$05.50$^{\prime \prime}$ & M\\
        J1350$-$2204 & 13h50m14.0902s & $-$22$^{\circ}$04$^{\prime}$41.0779$^{\prime \prime}$ & M\\
        J1358+4737 & 13h58m40.6651s & +47$^{\circ}$37$^{\prime}$58.317$^{\prime \prime}$ & M\\
        J1404$-$0130 & 14h04m45.8954s & $-$01$^{\circ}$30$^{\prime}$21.9472$^{\prime \prime}$ & M\\
        J1419$-$1928 & 14h19m49.7387s & $-$19$^{\circ}$28$^{\prime}$25.2679$^{\prime \prime}$ & M\\
        J1506$-$0919 & 15h06m03.035s & $-$09$^{\circ}$19$^{\prime}$12.054$^{\prime \prime}$ & M\\
        J1513+2338 & 15h13m40.1879s & +23$^{\circ}$38$^{\prime}$35.321$^{\prime \prime}$ & M\\
        J1555$-$2508 & 15h55m44.9838s & $-$25$^{\circ}$08$^{\prime}$11.875$^{\prime \prime}$ & M\\
        J1602+2418 & 16h02m13.841s & +24$^{\circ}$18$^{\prime}$37.838$^{\prime \prime}$ & M\\
        J1604$-$2223 & 16h04m01.4717s & $-$22$^{\circ}$23$^{\prime}$40.986$^{\prime \prime}$ & M\\
        J1735$-$0559 & 17h35m26.7845s & $-$05$^{\circ}$59$^{\prime}$50.215$^{\prime \prime}$ & M\\
        J1754+0459 & 17h54m17.520s & +04$^{\circ}$59$^{\prime}$39.60$^{\prime \prime}$ & M\\
        J1935$-$1602 & 19h35m35.7952s & $-$16$^{\circ}$02$^{\prime}$32.3744$^{\prime \prime}$ & M\\
        J2137+3455 & 21h37m44.1028s & +34$^{\circ}$55$^{\prime}$42.0919$^{\prime \prime}$ & M\\
        J2137$-$2042\tablenotemark{c,f} & 21h37m50.0079s & $-$20$^{\circ}$42$^{\prime}$31.6724$^{\prime \prime}$& S\\
        J2212+0152 & 22h12m37.9734s & +01$^{\circ}$52$^{\prime}$51.1855$^{\prime \prime}$ & M\\
        J2253+0236 & 22h53m21.1045s & +02$^{\circ}$36$^{\prime}$13.0405$^{\prime \prime}$ & M\\
        J2332+4030 & 23h32m52.9314s & +40$^{\circ}$30$^{\prime}$37.1361$^{\prime \prime}$ & M
\label{rejects}
\enddata
\tablenotetext{a}{15 GHz Stokes I pixels are 0.11 mas}
\tablenotetext{b}{Confirmed MSO}
\tablenotetext{c}{Indeterminate MSO}
\tablenotetext{d}{15 GHz Stokes I pixels are 0.2 mas}
\tablenotetext{e}{8 and 15 GHz Stokes I and 8-15 GHz spectral index map pixels are 0.23 mas}
\tablenotetext{f}{15 GHz Stokes I pixels are 0.18 mas}
\tablecomments{$^{\dag}$J0048+3157 was originally rejected in K24a. Pixel sizes for the Stokes I maps are 0.25 mas at 5 GHz, 0.2 mas at 8 GHz, and 0.1 mas at 15 GHz unless otherwise stated. Rejection Criterion key: M - morphology not consistent with a CSO, V - fractional variability in excess of 20\% yr$^{-1}$, S - size greater than 1 kpc}
\end{deluxetable}
\newpage

\section{Indeterminate A-Candidates} \label{indetsection}
\setcounter{table}{0}
This section contains a list of the 46 indeterminate A-candidates as well as their VLBA peak fluxes and flux densities.
\startlongtable
\begin{deluxetable}{ccc|cc|cc|cc}
\tablecaption{Indeterminate A-Candidate Sources}
    \tablehead{\colhead{Source Name} & \colhead{RA} & \colhead{Dec} & \colhead{$S_{5}$ [Jy]} & \parbox[c]{2cm}{\centering $S_{peak,5}$ [${\rm Jy\,beam}^{-1}$]} & \colhead{$S_{8}$ [Jy]} & \parbox[c]{2cm}{\centering $S_{peak,8}$ [${\rm Jy\,beam}^{-1}$]} & \colhead{$S_{15}$ [Jy]} & \parbox[c]{2cm}{\centering $S_{peak,15}$ [${\rm Jy\,beam}^{-1}$]}}
    \startdata
        J0146+2110 & 01h46m58.7838s & +21$^{\circ}$10$^{\prime}$24.3842$^{\prime \prime}$ & 0.663(33) & 0.169(8) & 0.273(14) & 0.063(3) & ... & ... \\
        J0210+0419 & 02h10m44.5136s & +04$^{\circ}$19$^{\prime}$34.8829$^{\prime \prime}$ & 0.339(17) & 0.079(4) & 0.142(7) & 0.029(1) & ... & ... \\
        J0234+0446 & 02h34m07.1554s & +04$^{\circ}$46$^{\prime}$43.091$^{\prime \prime}$ & 0.191(10) & 0.113(6) & 0.106(5) & 0.048(2) & ... & ... \\
        J0235$-$0100 & 02h35m16.810s & $-$01$^{\circ}$00$^{\prime}$52.00$^{\prime \prime}$ & 0.091(5) & 0.014(1) & 0.038(2) & 0.014(1) & ... & ... \\
        J0300+8202\tablenotemark{a} & 03h00m11.1379s & +82$^{\circ}$02$^{\prime}$39.3552$^{\prime \prime}$ & 0.534(27) & 0.079(4) & 0.184(9) & 0.026(1) & ... & ... \\
        J0301+3512 & 03h01m42.3291s & +35$^{\circ}$12$^{\prime}$20.3012$^{\prime \prime}$ & 0.009(0) & 0.008(0) & 0.015(1) & 0.011(1) & ... & ... \\
        J0407$-$2757 & 04h07m57.9308s & $-$27$^{\circ}$57$^{\prime}$05.4206$^{\prime \prime}$ & 0.595(30) & 0.154(8) & 0.218(11) & 0.042(2) & 0.088(4) & 0.020(1) \\
        J0457$-$0849 & 04h57m20.2128s & $-$08$^{\circ}$49$^{\prime}$05.484$^{\prime \prime}$ & 0.336(17) & 0.187(9) & 0.210(10) & 0.097(5) & 0.113(6) & 0.034(2) \\
        J0503+0203 & 05h03m21.1972s & +02$^{\circ}$03$^{\prime}$04.6769$^{\prime \prime}$ & 3.178(159) & 1.120(56) & 1.848(92) & 0.663(33) & 1.015(51) & 0.369(18) \\
        J0518+4730 & 05h18m12.0899s & +47$^{\circ}$30$^{\prime}$55.5282$^{\prime \prime}$ & 0.502(25) & 0.242(12) & 0.375(19) & 0.153(8) & 0.219(11) & 0.068(3) \\
        J0731$-$2224 & 07h31m31.5084s & $-$22$^{\circ}$24$^{\prime}$20.867$^{\prime \prime}$ & 0.939(47) & 0.392(20) & 0.655(33) & 0.167(8) & 0.334(17) & 0.051(3) \\
        J0733+5605 & 07h33m28.6148s & +56$^{\circ}$05$^{\prime}$41.73$^{\prime \prime}$ & 0.075(4) & 0.009(0) & 0.057(3) & 0.012(1) & ... & ... \\
        J0756+6347 & 07h56m54.6107s & +63$^{\circ}$47$^{\prime}$59.022$^{\prime \prime}$ & 0.333(17) & 0.259(13) & 0.243(12) & 0.164(8) & 0.147(7) & 0.061(3) \\
        J0811+4308 & 08h11m37.364s & +43$^{\circ}$08$^{\prime}$29.422$^{\prime \prime}$ & 0.183(9) & 0.083(4) & 0.116(6) & 0.047(2) & ... & ... \\
        J0846$-$2610 & 08h46m00.7338s & $-$26$^{\circ}$10$^{\prime}$54.155$^{\prime \prime}$ & 0.531(27) & 0.087(4) & 0.255(13) & 0.033(2) & ... & ... \\
        J0853+6722 & 08h53m34.3233s & +67$^{\circ}$22$^{\prime}$15.6614$^{\prime \prime}$ & 0.307(15) & 0.131(7) & 0.177(9) & 0.063(3) & 0.075(4) & 0.024(1) \\
        J0934+4908 & 09h34m15.7652s & +49$^{\circ}$08$^{\prime}$21.718$^{\prime \prime}$ & 0.403(20) & 0.241(12) & 0.272(14) & 0.104(5) & 0.162(8) & 0.050(2) \\
        J0943$-$0819 & 09h43m36.9446s & $-$08$^{\circ}$19$^{\prime}$30.8192$^{\prime \prime}$ & 0.474(24) & 0.074(4) & 0.293(15) & 0.039(2) & ... & ... \\
        J0945+2729 & 09h45m15.6246s & +27$^{\circ}$29$^{\prime}$11.351$^{\prime \prime}$ & 0.155(8) & 0.066(3) & 0.123(6) & 0.042(2) & ... & ... \\
        J1031$-$2228 & 10h31m52.3121s & $-$22$^{\circ}$28$^{\prime}$24.974$^{\prime \prime}$ & 0.226(11) & 0.070(4) & 0.221(11) & 0.075(4) & 0.118(6) & 0.039(2) \\
        J1057+0012 & 10h57m15.7674s & +00$^{\circ}$12$^{\prime}$03.575$^{\prime \prime}$ & 0.283(14) & 0.115(6) & 0.201(10) & 0.060(3) & 0.098(5) & 0.024(1) \\
        J1109+1043 & 11h09m46.0687s & +10$^{\circ}$43$^{\prime}$43.4606$^{\prime \prime}$ & 0.311(16) & 0.057(3) & 0.177(9) & 0.028(1) & 0.043(2) & 0.027(1) \\
        J1133+7831 & 11h33m59.8007s & +78$^{\circ}$31$^{\prime}$22.4269$^{\prime \prime}$ & 0.056(3) & 0.012(1) & 0.030(1) & 0.012(1) & ... & ... \\
        J1135$-$0021 & 11h35m13.0119s & $-$00$^{\circ}$21$^{\prime}$18.9813$^{\prime \prime}$ & 0.226(11) & 0.062(3) & 0.098(5) & 0.031(2) & ... & ... \\
        J1227+4400 & 12h27m41.9842s & +44$^{\circ}$00$^{\prime}$42.0804$^{\prime \prime}$ & 0.010(1) & 0.009(0) & 0.019(1) & 0.019(1) & 0.027(1) & 0.025(1) \\
        J1241+5458 & 12h41m27.703s & +54$^{\circ}$58$^{\prime}$19.063$^{\prime \prime}$ & 0.139(7) & 0.062(3) & 0.084(4) & 0.028(1) & ... & ... \\
        J1247+2551 & 12h47m44.5375s & +25$^{\circ}$51$^{\prime}$55.352$^{\prime \prime}$ & 0.077(4) & 0.031(2) & 0.050(3) & 0.015(1) & ... & ... \\
        J1251+2102 & 12h51m27.7008s & +21$^{\circ}$02$^{\prime}$53.639$^{\prime \prime}$ & 0.113(6) & 0.063(3) & 0.079(4) & 0.035(2) & ... & ... \\
        J1307+7649 & 13h07m05.2451s & +76$^{\circ}$49$^{\prime}$18.1545$^{\prime \prime}$ & 0.161(8) & 0.029(1) & 0.093(5) & 0.014(1) & ... & ... \\
        J1311+1417 & 13h11m07.8242s & +14$^{\circ}$17$^{\prime}$46.648$^{\prime \prime}$ & 0.374(19) & 0.161(8) & 0.239(12) & 0.076(4) & 0.112(6) & 0.030(1) \\
        J1317+4115 & 13h17m39.1954s & +41$^{\circ}$15$^{\prime}$45.621$^{\prime \prime}$ & 0.222(11) & 0.074(4) & 0.167(8) & 0.041(2) & ... & ... \\
        J1322+2645\tablenotemark{b} & 13h22m14.9729s & +26$^{\circ}$45$^{\prime}$46.355$^{\prime \prime}$ & 0.155(8) & 0.014(1) & 0.130(7) & 0.008(0) & ... & ... \\
        J1325+2109 & 13h25m18.7089s & +21$^{\circ}$09$^{\prime}$25.281$^{\prime \prime}$ & 0.120(6) & 0.060(3) & 0.069(3) & 0.026(1) & ... & ... \\
        J1357+4353 & 13h57m40.587s & +43$^{\circ}$53$^{\prime}$59.772$^{\prime \prime}$ & 0.378(19) & 0.147(7) & 0.228(11) & 0.054(3) & ... & ... \\
        J1409$-$2315 & 14h09m11.97s & $-$23$^{\circ}$15$^{\prime}$49.5$^{\prime \prime}$ & 0.219(11) & 0.082(4) & 0.193(10) & 0.048(2) & ... & ... \\
        J1421+7513 & 14h21m15.0189s & +75$^{\circ}$13$^{\prime}$20.2578$^{\prime \prime}$ & 0.074(4) & 0.019(1) & 0.043(2) & 0.019(1) & 0.025(1) & 0.021(1) \\
        J1435+7605 & 14h35m47.0981s & +76$^{\circ}$05$^{\prime}$25.8231$^{\prime \prime}$ & 0.498(25) & 0.049(2) & 0.287(14) & 0.012(1) & ... & ... \\
        J1543$-$0757 & 15h43m01.6875s & $-$07$^{\circ}$57$^{\prime}$06.629$^{\prime \prime}$ & 0.828(41) & 0.171(9) & 0.653(33) & 0.062(3) & ... & ... \\
        J1559+1624 & 15h59m25.0704s & +16$^{\circ}$24$^{\prime}$40.895$^{\prime \prime}$ & 0.193(10) & 0.081(4) & 0.114(6) & 0.038(2) & 0.061(3) & 0.021(1) \\
        J1642+6655 & 16h42m21.9273s & +66$^{\circ}$55$^{\prime}$49.4917$^{\prime \prime}$ & 0.066(3) & 0.025(1) & 0.042(2) & 0.014(1) & 0.034(2) & 0.033(2) \\
        J1753+2750 & 17h53m01.3459s & +27$^{\circ}$50$^{\prime}$59.0172$^{\prime \prime}$ & 0.264(13) & 0.171(9) & 0.142(7) & 0.074(4) & 0.065(3) & 0.021(1) \\
        J2014+5059 & 20h14m28.59s & +50$^{\circ}$59$^{\prime}$09.5286$^{\prime \prime}$ & 0.316(16) & 0.071(4) & 0.111(6) & 0.028(1) & ... & ... \\
        J2058+0542\tablenotemark{b} & 20h58m28.875s & +05$^{\circ}$42$^{\prime}$51.0133$^{\prime \prime}$ & 0.305(15) & 0.050(2) & 0.109(5) & 0.015(1) & ... & ... \\
        J2244+2600 & 22h44m35.1473s & +26$^{\circ}$00$^{\prime}$20.702$^{\prime \prime}$ & 0.359(18) & 0.255(13) & 0.216(11) & 0.106(5) & 0.125(6) & 0.045(2) \\
        J2248+7718\tablenotemark{c} & 22h48m34.6926s & +77$^{\circ}$18$^{\prime}$51.9528$^{\prime \prime}$ & 0.088(4) & 0.019(1) & 0.033(2) & 0.011(1) & ... & ... \\
        J2355$-$2125 & 23h55m02.1466s & $-$21$^{\circ}$25$^{\prime}$36.7821$^{\prime \prime}$ & 0.478(24) & 0.235(12) & 0.292(15) & 0.124(6) & 0.129(6) & 0.039(2)
    \enddata
    \tablenotetext{a}{5 GHz Stokes I, 5-8 GHz spectral index, and 5-15 GHz spectral index pixels are 0.3 mas, 8 GHz Stokes I and 8-15 spectral index pixels are 0.26 mas, 15 GHz Stokes I pixels are 0.13 mas}
    \tablenotetext{b}{5 GHz Stokes I, 5-8 GHz spectral index, and 5-15 GHz spectral index pixels are 0.35 mas, 15 GHz Stokes I pixels are 0.15 mas}
    \tablenotetext{c}{8 GHz Stokes I and 8-15 spectral index pixels are 0.22 mas, 15 GHz Stokes I pixels are 0.11 mas}
    \tablecomments{No 15 GHz flux values indicates the peak flux was below our 20 m${\rm Jy\,beam}^{-1}$ cutoff. Pixel sizes for the Stokes I maps are 0.25 mas at 5 GHz, 0.2 mas at 8 GHz, and 0.1 mas at 15 GHz unless otherwise stated.}
    \label{indets}
\end{deluxetable}

\end{document}